\documentclass[aps,prd,floatfix,showpacs,superscriptaddress,nofootinbib,notitlepage]{revtex4-1}

\usepackage{dsfont}

\usepackage{amsmath}

\usepackage{amsfonts}

\usepackage{amssymb}

\usepackage{graphicx}%

\usepackage{subfig}

\usepackage{braket}

\usepackage{float}

\usepackage{slashed}

\usepackage{color}

\usepackage{tensor}

\definecolor{darkgreen}{rgb}{0,0.35,0}
\usepackage[colorlinks=true, pdfstartview=FitV, linkcolor=blue, citecolor=red, urlcolor=magenta]{hyperref}

\def\m{\mu}
\def\n{\nu}

\def\l{\lambda}

\def\r{\rho}
\def\s{\sigma}

\newcommand{\MSbar}{\overline{\mbox{MS}}}

\newcommand{\be}{\begin{equation}}
\newcommand{\ee}{\end{equation}}

\newcommand{\beq}{\begin{eqnarray}}
\newcommand{\eeq}{\end{eqnarray}}

\newcommand{\kulak}{KU Leuven Campus Kortrijk---Kulak, Department of Physics, Etienne Sabbelaan 53 bus 7657, 8500 Kortrijk, Belgium}
\newcommand{\ughent}{Ghent University, Department of Physics and Astronomy, Krijgslaan 281-S9, 9000 Gent, Belgium}
\newcommand{\uerj}
{Universidade do Estado do Rio de Janeiro,
	Instituto de F\'isica---Departamento de F\'isica Te\'orica---Rua S\~{a}o Francisco Xavier 524,
	20550-013, Maracan\~{a}, Rio de Janeiro, Brasil}

\begin{document}	
	\title{ Spectral properties of local gauge invariant composite operators in the $SU(2)$ Yang--Mills--Higgs model}
	\author{D.~Dudal}\email{david.dudal@kuleuven.be}\affiliation{\kulak}\affiliation{\ughent}
	\author{D.~M.~van Egmond }\email{duifjemaria@gmail.com}\affiliation{\uerj}
	\author{M.~S.~Guimar\~{a}es}\email{msguimaraes@uerj.br}\affiliation{\uerj}
	\author{L.~F.~Palhares}\email{leticia.palhares@uerj.br}\affiliation{\uerj}
	\author{G.~Peruzzo}\email{gperuzzofisica@gmail.com}\affiliation{\uerj}
	\author{S.~P.~Sorella}\email{silvio.sorella@gmail.com}\affiliation{\uerj}
	\begin{abstract}
		The spectral properties of a set of local gauge (BRST) invariant composite operators are investigated in
		the $SU(2)$ Yang--Mills--Higgs model with a single Higgs field in the fundamental representation, quantized in the 't Hooft $R_{\xi}$-gauge. These operators can be thought of as  a
		BRST invariant version of the elementary fields of the theory, the Higgs and  gauge fields, with  which they share a gauge independent pole mass. The two-point correlation functions of both  BRST invariant composite operators and  elementary fields, as well as their spectral functions, are investigated at one-loop order. It is shown that the spectral functions of the elementary fields suffer from a strong unphysical dependence from  the gauge parameter $\xi$, and can even exhibit positivity violating behaviour. In contrast, the BRST invariant local operators exhibit a well defined positive spectral density.

	\end{abstract}
	
	\maketitle
	\section{Introduction}
	
	The principle of  gauge invariance is the  ultimate guideline to formulate quantum field theories of the fundamental interactions as, for example, the electroweak theory \cite{tHooft:1980xss,Peskin:1995ev}. In non-Abelian gauge theories, genuine local gauge invariant quantities are associated to  composite operators. It is therefore remarkable that  the Standard Model is successfully described by employing non-gauge invariant fields as the Higgs and the $W$ and $Z$  elementary fields. Needless to say, the high order calculations of the pole masses and cross sections worked out by means of these non-gauge invariant fields are in very accurate agreement with the experimental data,  see e.g.~\cite{Jegerlehner:2001fb,Jegerlehner:2002em,Martin:2015lxa,Martin:2015rea} for a few illustrations.
	
	At the theoretical level, the gauge parameter independence of the pole masses  of both  transverse $W$ and $Z$ bosons as well as of the Higgs field two-point correlation functions are understood by means of the so-called  Nielsen identities \cite{Nielsen:1975fs,Aitchison:1983ns,Piguet:1984js,gambino1999fermion,gambino2000nielsen,Grassi:2000dz,Andreassen:2014eha}, which follow from the Slavnov-Taylor identities encoding the exact BRST symmetry of quantized non-Abelian gauge theories. Nevertheless, as one easily  figures out, the direct use of the non-gauge invariant fields displays several limitations, which become more severe in the case of a non-Abelian gauge theory. For instance, in the case of the $U(1)$ Higgs model, the transverse component of the Abelian gauge field $A_\mu$ is gauge invariant, so that the two-point correlation function
	$\mathcal{P}_{\mu\nu}(p) \langle A_\mu(p) A_\nu(-p) \rangle$, where  $\mathcal{P}_{\mu\nu}(p) = (\delta_{\mu\nu} -\frac{p_\mu p_\nu}{p^2})$ is the transverse  projector, turns out to be independent from the gauge parameter $\xi$. However, this is no more true in the non-Abelian case, where both Higgs and gauge boson two-point functions, {\it i.e.}~$\langle h(p) h(-p)$ and $\mathcal{P}_{\mu\nu}(p) \langle A^a_\mu(p) A^b_\nu(-p) \rangle$, where $h$ stands for the Higgs field and $A^a_\mu$ for the gauge boson field,  exhibit a strong gauge dependence from $\xi$. As a consequence, the understanding of the two-point  correlation functions of both Higgs field $h$ and gauge vector boson $A^a_\mu$ in terms of the K\"all\'en-Lehmann (KL) spectral representation is completely jeopardized by an unphysical dependence from the gauge parameter $\xi$, obscuring a direct interpretation of the above mentioned correlation functions in terms of the elementary excitations of the physical spectrum, namely the Higgs and the vector gauge boson particles.  We also note here that from a lattice perspective, it is expected that the spectrum of a gauge (Higgs) theory should be describable in terms of local gauge invariant operator correlation functions, with concrete physical information hiding in the various (positive and gauge invariant) spectral functions, not only pole masses, decay widths, but also transport coefficients at finite temperature etc. Clearly, such information will not correctly be encoded in gauge variant, non-positive spectral functions.

	Within this perspective, the use of manifest gauge invariant variables to describe the Higgs and the vector gauge bosons is certainly very welcome. This endeavour was  first proposed by 't Hooft in \cite{tHooft:1980xss}, and later on formalized by Fr\"ohlich, Morchio and Strocchi (FMS) in \cite{Frohlich:1980gj, Frohlich:1981yi}. These authors were able to build, out of the elementary fields, a set of local composite  gauge invariant operators $\{\tilde{\mathcal{O}}(x)\}$ which, when expanded around the value
	$\Phi=  constant$ which minimizes the Higgs potential present in the starting classical action, give rise to two-point functions which enjoy the important property of reproducing, at the tree level, the two-point correlation functions of the elementary fields $\{ \varphi \}= (A^a_\mu,h)$, namely
	\begin{equation}
	\braket{\tilde{\mathcal{O}}(x) \tilde {\mathcal{O}}(y)}\sim    \braket{\varphi(x) \varphi(y)}_{\rm tree}+ \ldots,
	\label{1a}
	\end{equation}
	where $ \ldots$ denote the higher order loop corrections which will be the main subject of the present
	work. Equation \eqref{1a} shows in a very simple and intuitive way the relevance of the composite operators $\{\tilde {\mathcal{O}}(x)\}$ in order to provide a description of the gauge vector bosons and of the Higgs particle within a fully gauge invariant environment, see also the recent works  \cite{maas2015field, Maas:2017wzi,Maas:2018xxu,Sondenheimer:2019idq} where, amongst other things, a lattice formulation has been proposed. Certain aspects of a gauge invariant version of the Higgs phenomenon were also covered in \cite{Kondo:2016ywd,Kondo:2018qus}, albeit whilst assuming the ``frozen'' radial limit, $\varphi^\dagger \varphi=\textrm{fixed}$, corresponding to a Higgs coupling $\lambda\to \infty$, a formal limit hampering explicit computations in the continuum.
	
	In two earlier works \cite{Dudal:2019aew, Dudal:2019pyg}, we have laid the ground for the study of the spectral properties of the gauge (BRST) invariant local composite operators $\{\tilde {\mathcal{O}}(x)\}$ in the FMS framework. In \cite{Dudal:2019pyg}, we have made the first analytic one-loop calculations of these BRST invariant
	operators in the simpler $U(1)$ Higgs model quantized in the $R_{\xi}$-gauge. In particular, we have worked out the one-loop corrections to the two-point functions in eq.~\eqref{1a} corresponding to the Higgs and Abelian gauge fields and we have shown that they have the same gauge independent pole masses of the corresponding elementary two-point correlation functions. In addition, we have explicitly shown that the correlation functions of the  composite operators display a well defined positive and gauge independent K\"all\'en-Lehmann  spectral representation, a feature not shared by the two-point correlation functions of the elementary fields which, as in the explicit case of the Higgs field, {\it i.e.}~$\langle h(p) h(-p) \rangle$, display an  unphysical dependence from the gauge parameter $\xi$, becoming even negative depending on the value of $\xi$.  Moreover, in \cite{Capri:2020ppe}, the renormalization properties of these composite operators were scrutinized using the algebraic renormalization approach.
	
	The aim of the present work is that of extending the techniques of \cite{Dudal:2019aew, Dudal:2019pyg} to the more complex case of $SU(2)$ Higgs model with a single Higgs field in the fundamental representation. As we shall see, besides the exact BRST invariance, the quantized theory exhibits a global $SU(2)$ symmetry  commonly referred to as the  {\it{custodial symmetry}}. Moreover, the local composite BRST invariant operators corresponding to the gauge bosons  transform as a triplet under the custodial symmetry, a property which will imply useful relations for their two-point correlation functions.
	
	The present work is organized as follows. In section~\ref{I}, we give a review of the $SU(2)$ Yang--Mills--Higgs model with a single Higgs field in the fundamental representation, of the gauge fixing procedure and its ensuing BRST invariance.  In section~\ref{III} we calculate the two-point correlation functions of the elementary fields up to one-loop order. In section~\ref{IIIIr}, we define the BRST invariant local composite operators  $(O(x), R^a_\mu(x))$ corresponding to the BRST invariant extension of $(h,A^a_\mu)$ and calculate their one-loop correlation functions. In section~\ref{V}, we discuss the spectral properties of both  elementary and composite operators.
	
	In order to give a more general idea of the  behavior of the spectral functions, we shall be using two sets of parameters which we shall refer as to Region I and Region II. To some extent, Region II can be associated to the perturbative weak coupling regime, while in Region I we keep the gauge coupling a little bit larger,  while decreasing the vev  (vacuum expectation value) of the Higgs field. Section~\ref{VI} is devoted to our conclusion and outlook. The technical details are all collected in the Appendices.

	\section{The action and its symmetries \label{I}}
	The Yang--Mills action with a single Higgs field in the fundamental representation is given by
	\begin{eqnarray}
	S_0&=& \int d^4x \left\{ \frac{1}{4}F_{\mu\nu}^a F_{\mu\nu}^a+(D_{\mu}^{ij}\Phi^{\dagger j})(D_{\mu}^{ik}\Phi^k)+\frac{\lambda}{2}(\Phi^{\dagger i}\Phi^i-\frac{1}{2}v^2)^2\right\}\nonumber \\
	&=& S_{\rm YM}+S_{\rm Higgs} \label{1}
	\end{eqnarray}
	with
	\begin{eqnarray}
	F_{\mu\nu}=\partial_{\mu}A_{\nu}^a-\partial_{\nu}A_{\mu}^a+g \epsilon^{abc}A_{\mu}^bA_{\nu}^c
	\end{eqnarray}
	and
	\begin{eqnarray}
	D_{\mu}^{ij}\Phi^j=\partial_{\mu}\Phi^i-\frac{i}{2}g(\tau^a)^{ij}A_{\mu}^a\Phi^j, \,\,\, (D_{\mu}^{ij}\Phi^j)^{\dagger}=\partial_{\mu}\Phi^{i\dagger}+\frac{i}{2}g\Phi^{j\dagger}(\tau^a)^{ji}A_{\mu}^a,
	\end{eqnarray}
	with the Pauli matrices $\tau^a (a=1,2,3)$ and the Levi-Civita tensor  $\epsilon^{abc}$ referring to the gauge symmetry group $SU(2)$. The scalar complex field $\Phi^i(x)$ is in the fundamental representation of $SU(2)$,
	{\it i.e.}~$i,j=1,2$. Thus, $\Phi$ is an $SU(2)$-doublet of complex scalar fields that can be written as
	\beq
	\Phi&=&\frac{1}{\sqrt{2}}\begin{pmatrix}
		\phi^+ \\
		\phi^0
	\end{pmatrix}=\frac{1}{\sqrt{2}}\begin{pmatrix}
		\phi_1+i \phi_2 \\
		\phi_3+i \phi_4
	\end{pmatrix}.
	\eeq
	The configuration which minimizes the Higgs potential in the expression \eqref{1}  is
	\begin{eqnarray}
	\braket{\Phi}&=\frac{1}{\sqrt{2}}
	\begin{pmatrix}
	v \\
	0 \end{pmatrix} \label{aa}
	\end{eqnarray}
	and we write down  $\Phi(x) $ as an expansion  around the configuration \eqref{aa}, so that
	\begin{eqnarray}
	\Phi= \frac{1}{\sqrt{2}}
	\begin{pmatrix}
	v+h+i\rho_3 \\
	i\rho_1-\rho_2\end{pmatrix}, \label{expansion1}
	\end{eqnarray}
	where $h$ is the Higgs field and $\rho^a$, $a=1,2,3$, the would-be Goldstone bosons. We can use the  matrix notation\footnote{ This is of course possible thanks to the fact that $\Phi$ counts 3 Goldstone modes and that $SU(2)$ has three generators. This ``numerology'' is essentially what leads to a large custodial symmetry in the $SU(2)$ case. }:
	\begin{eqnarray}
	\Phi=\frac{1}{\sqrt{2}}((v+h)\textbf{1}+i\rho^a \tau^a) \cdot
	\begin{pmatrix}
	1\\
	0
	\end{pmatrix}, \label{expansion2}
	\end{eqnarray}
	so that the second term in eq.~\eqref{1} becomes
	\begin{eqnarray}
	(D^{ij}_{\mu}\Phi^j)^{\dagger}D_{\mu}^{ik}\Phi^k &=& \frac{1}{2} (1,0) \cdot \Bigg[\partial_{\mu}h \cdot \textbf{1}-i\partial_{\mu}\rho^a \tau^a+\frac{ig}{2}\tau^a A_{\mu}^a \Big((v+h)\textbf{1}-i\rho^b\tau^b\Big)\Bigg]\\
	&\times&\Bigg[\partial_{\mu}h \cdot \textbf{1}+i\partial_{\mu}\rho^c \tau^c-\frac{ig}{2} ((v+h)\textbf{1}\nonumber-i\rho^d\tau^d\Big)\tau^c A_{\mu}^c\Bigg] \cdot \begin{pmatrix} 1\\
	0 \end{pmatrix}\nonumber\\
	&=& \tilde{\mathcal{L}_0}+ \tilde{\mathcal{L}_1}+ \tilde{\mathcal{L}_2},
	\end{eqnarray}
	with $\tilde{\mathcal{L}_i}$ the $i$th term in powers of $A_{\mu}$:
	\begin{eqnarray}
	\tilde{\mathcal{L}}_0&=&\frac{1}{2}\left((\partial_{\mu}h)^2+\partial_{\mu}\rho^a\partial_{\mu}\rho^a\right), \nonumber\\
	\tilde{\mathcal{L}}_1&=&-\frac{1}{2}\left\{gv A_{\mu}^a\partial_{\mu} \rho^a-gA_{\mu}^a\rho^a\partial_{\mu}h+gA_{\mu}^a(\partial_{\mu}\rho^a)h+g \epsilon^{abc} \partial_{\mu} \rho^a \rho^b A_{\mu}^c \right\},\nonumber\\
	\tilde{\mathcal{L}_2}&=&\frac{g^2}{8}A_{\mu}^a A_{\mu}^a \left[(v+h)^2+\rho^b\rho^b \right],
	\end{eqnarray}
	and we have the full action
	\begin{eqnarray}
	S_0&=&\int d^4x \frac{1}{2}\Bigg\{\frac{1}{2}F^a_{\mu\nu}F^a_{\mu\nu}+\frac{1}{4}v^2g^2A_{\mu}^aA_{\mu}^a+(\partial_{\mu}h)^2+\partial_{\mu}\rho^a\partial_{\mu}\rho^a-gvA_{\mu}^a\partial_{\mu}\rho^a+gA_{\mu}^a\rho^a\partial_{\mu}h-gA^a_{\mu}(\partial_{\mu}\rho^a)h\nonumber\\
	&-&g \epsilon^{abc}\partial_{\mu}\rho^a\rho^b A_{\mu}^c+\frac{g^2}{4}A_{\mu}^a A_{\mu}^a\left[2vh+h^2+\rho^b\rho^b\right]+\lambda v^2h^2+\lambda v h (h^2+\rho^a \rho^a)\nonumber\\
	&+&\frac{\lambda}{4}(h^2+\rho^a\rho^a)^2 \Bigg\}.
	\end{eqnarray}
	One sees that both gauge field $A^a_\mu$ and Higgs field $h$ have acquired a mass given, respectively, by
	\beq
	m^2= \frac{1}{4} g^2 v^2, \,\,\,\,\,\,\, m_h^2= \lambda v^2 \;.
	\eeq
	\subsection{Gauge fixing and BRST symmetry}
	The action \eqref{1} is invariant under the local $\omega$-parametrized gauge transformations
	\begin{equation}
	\delta A_{\mu}^a=-D_{\mu}^{ab}\omega^b, \,\,\, \delta \Phi=-\frac{ig}{2}\omega^a\tau^a\Phi, \,\,\, \delta \Phi^{\dagger}=\frac{ig}{2}\omega^a \Phi^{\dagger}\tau^a,
	\end{equation}
	which, when written in terms of the fields $(h,\rho^a)$, become
	\begin{equation}
	\delta h = \frac{g}{2}\omega^a \rho^a, \,\,\, \delta \rho^a=-\frac{g}{2}(\omega^a(v+h)\textbf{1}-\epsilon^{abc}\omega^b\rho^c).
	\end{equation}
	As done in the $U(1)$ case  \cite{Dudal:2019aew, Dudal:2019pyg}, we shall be using the $R_{\xi}$-gauge. We add thus need the gauge fixing term
	\beq
	\mathcal{S}_{\rm gf}&=&s \int d^4 x  \Bigl\{-i \frac{\xi}{2}\bar{c}^a b^a+\bar{c}^a(\partial_{\m} A_{\m}^a-\xi m \rho^a)\Bigr\} \nonumber\\
	&=& \frac{1}{2}\int d^4 x \Bigl\{ \xi b^a b^a+ 2 i b^a \partial_{\m}  A_{\m}^a+2\bar{c}^a \partial_{\m}D_{\m}^{ab}c^b-2i \xi m b^a \rho^a \nonumber\\
	&-& 2 \xi \bar{c}^a m c^a - g \xi \bar{c}^a m h c^a - \xi g \epsilon^{abc} \bar{c}^a  c^b \rho^c\Bigr\} \;,
	\eeq
	so that the gauge fixed action $S_{\text{full}} = S_0 +\mathcal{S}_{\rm gf} $, namely
	\begin{eqnarray}
	S_{\text{full}}&=&\int d^4x \frac{1}{2}\Bigg\{\frac{1}{2}F^a_{\mu\nu}F^a_{\mu\nu}+\frac{1}{4}v^2g^2A_{\mu}^aA_{\mu}^a\nonumber\\&+&(\partial_{\mu}h)^2+\partial_{\mu}\rho^a\partial_{\mu}\rho^a-gvA_{\mu}^a\partial_{\mu}\rho^a+gA_{\mu}^a\rho^a\partial_{\mu}h-gA^a_{\mu}(\partial_{\mu}\rho^a)h\nonumber\\
	&-&g \epsilon^{abc}\partial_{\mu}\rho^a\rho^b A_{\mu}^c+\frac{g^2}{4}A_{\mu}^a A_{\mu}^a\left[2vh+h^2+\rho^b\rho^b\right]+\lambda v^2h^2\nonumber\\
	&+&\lambda v h (h^2+\rho^a \rho^a)+\frac{\lambda}{4}(h^2+\rho^a\rho^a)^2 +\xi b^a b^a+ 2 i b^a \partial_{\m}  A_{\m}^a+2\bar{c}^a \partial_{\m}D_{\m}^{ab}c^b-2i \xi m b^a \rho^a \nonumber\\
	&-& 2 \xi \bar{c}^a m c^a - g \xi \bar{c}^a m h c^a - \xi g \epsilon^{abc} \bar{c}^a  c^b \rho^c \Bigg\}
	\label{Sfull}
	\end{eqnarray}
	turns out to be left  invariant by the BRST transformations
	\begin{eqnarray}
	sA_{\mu}^a&=&-D_{\mu}^{ab}c^b, \,\,\, s h = \frac{g}{2}c^a \rho^a, \,\,\,  s \rho^a=-\frac{g}{2}(c^a(v+h)-\epsilon^{abc}c^b\rho^c)\nonumber\\
	sc^a&=&\frac{1}{2}g \epsilon^{abc}c^bc^c, \,\,\, s \bar{c}^a=ib^a, \,\,\, sb^a=0 \;,
	\end{eqnarray}
	\begin{equation}
	s S_{\text{full}} = 0 \;. \label{brstgt}
	\end{equation}	
	The Feynman rules for the full action \eqref{Sfull} are given in Appendix \ref{appA}. Notice that in the $R_{\xi}$-gauge the tree-level propagator $\braket{A^a_{\mu}(x) \rho^a(y)}$ vanishes, a well-known feature of this gauge choice \cite{Peskin:1995ev}. We will assume that $\xi\geq0$ to avoid tachyon poles in elementary propagators, see eq.~\eqref{props}.
	\subsection{Custodial symmetry \label{cust}}
	As already mentioned, apart from the BRST symmetry, there is an extra global symmetry, which we shall refer to as the custodial symmetry:
	\beq
	\delta A^a_\mu &=& \epsilon^{abc} \beta^b A^c_\mu, \nonumber\\
	\delta \rho^a &=& \epsilon^{abc} \beta^b \rho^c,                      \nonumber\\
	\delta \overline{c}^a&=& \epsilon^{abc} \beta^b \overline{c}^c, \nonumber
	\\
	\delta c^a&=& \epsilon^{abc} \beta^b c^c,
	\nonumber\\
	\delta b^a&=& \epsilon^{abc} \beta^b b^c, \nonumber\\
	\delta h&=&0 \;, \label{custodial}
	\eeq
	where $\beta^a$ is a constant parameter, $\partial_{\mu}\beta^a=0$,
	\begin{equation}
	\delta S_{\text{full}} = 0 \;. \label{beta}
	\end{equation}
	One notices that all fields carrying the index $a=1,2,3$, {\it i.e.}~$(A^a_\mu, b^a, c^a, {\bar c}^a, \rho^a)$, undergo a global transformation in the adjoint representation of $SU(2)$. The origin of this symmetry is an $SU(2)_{\rm gauge} \times SU(2)_{\rm global}$ symmetry of the action in the unbroken phase, see Appendix \ref{appcust}. The exception is the Higgs field $h$, which is left invariant, {\it i.e.}~it is a singlet. As we shall see in the following, this additional global symmetry will provide useful relationships for the two-point correlation functions of the BRST invariant composite operators.
		\section{One-loop evaluation of the	correlation function of the elementary fields \label{III}}
	For the elementary fields $h(x)$ and $A^a_{\mu}$, the correlation functions are calculated up to first loop order in Appendices \ref{hp} and \ref{Ap}. In what follows, we will always spell out again the momentum-dependent logarithms and explicit Feynman parameter dependence, that is, we will in the eventual correlation functions replace again the notational shorthands introduced in eq.~\eqref{eta}. The Feynman parameter integration itself was handled via \eqref{bigI}.  We have used the explicit expression \eqref{bigI} to numerically construct our spectral function plots (see later). More about that integral \eqref{bigI} can be found in \cite{tHooft:1978jhc,Ellis:2007qk}.
	
	For the Higgs field, for the propagator we get
	\beq
	\braket{h(p)h(-p)}&=&\frac{1}{p^2+m_h^2}+\frac{1}{(p^2+m_h^2)^2} \Pi_{hh}(p^2)+\mathcal{O}(\hbar^2),
	\label{hh}
	\eeq
	with $\Pi_{hh}(p^2)$ the one-loop correction to the self-energy calculated in Appendix \ref{hp}. For $d=4$, this correction is divergent. Employing the procedure of dimensional regularization, {\it i.e.}~setting $d=4-\epsilon$, the  divergent part for $\Pi_{hh}(p^2)$ is given by:
	\beq
	\Pi_{hh, \rm div}(p^2)&=&\frac{g^2 \left(\frac{3 m_h^4}{m^2}-3 \xi  m_h^2-3 \xi  p^2+9 p^2\right)}{32 \pi ^2 \epsilon },
	\label{hhdiv}
	\eeq
	which, following the  $\MSbar$-scheme, is re-absorbed by the introduction of suitable local  counterterms. We remain thus with the finite part of the Higgs self-energy
	\beq
	\Pi_{hh}(p^2)&=&\frac{3g^2}{{8 (4 \pi )^2}}\int_0^1 dx \, \Bigg\{2  \xi  \left(m_h^2+p^2\right) \ln \left(\frac{m^2 \xi }{\mu ^2}\right)-2 \xi  m_h^2+2 \left(6 m^2-p^2\right) \ln \left(\frac{m^2}{\mu ^2}\right)\nonumber\\
	&-&(12 m^2+\frac{p^4}{m^2}+4 p^2) \ln \left(\frac{m^2+p^2 (1-x) x}{\mu ^2}\right)+\left(\frac{p^4 }{m^2}-\frac{m_h^4}{m^2}\right)\ln \left(\frac{m^2 \xi +p^2 (1-x) x}{\mu ^2}\right)-12 m^2-2 \xi  p^2+2 p^2\nonumber\\
	&-&\frac{ m_h^4}{m^2} \left(-2 \ln \left(\frac{m_h^2}{\mu ^2}\right)+3 \ln \left(\frac{m_h^2+p^2 (1-x) x}{\mu ^2}\right)+2\right)\Bigg\}.
	\label{hhf}
	\eeq
	Before trying to resum the self-energy $\Pi_{hh}(p^2) $, we notice that this resummation is tacitly assuming that the
	second term in \eqref{hh} is much smaller than the first term. However, we see that eq.~\eqref{hh} contains terms of the order of $ \frac{p^4}{(p^2+m_h^2)^2}\ln \left(\frac{m^2+p^2 (1-x) x}{\mu ^2}\right)$  which cannot be resummed for big values of $p^2$.
	
	Indeed, if one proceeds naively and include these large contributions into a resummation, spurious tachyon poles will be induced in the correlator, analogous as in the earlier investigated $U(1)$ case in \cite{Dudal:2019pyg}. The need for care in resumming these contributions was also strengthened and worked out in great detail in \cite{Maas:2020kda}.
	
	We therefore proceed as in \cite{Dudal:2019aew, Dudal:2019pyg} and use the identity
	\beq
	p^4=(p^2+m_h^2)^2-m_h^4-2 p^2 m_h^2
	\eeq
	to rewrite
	\beq
	\frac{p^4}{(p^2+m_h^2)^2} \ln \frac{p^2x(1-x)+m^2}{\mu^2} &=& \ln \frac{p^2x(1-x)+m^2}{\mu^2} -\underline{\frac{(m_h^4+2p^2m_h^2)}{(p^2+m_h^2)^2} \ln \frac{p^2x(1-x)+m^2}{\mu^2}}.
	\label{jju}
	\eeq
	The term which has been underlined  in eq.~\eqref{jju} can be safely resummed, as it decays fast enough for large values of $p^2$.  We thence rewrite
	\beq
	\frac{{\Pi}_{hh}(p^2)}{(p^2+m_h^2)^2}&=& \frac{\hat{\Pi}_{hh}(p^2)}{(p^2+m_h^2)^2}+C_{hh}(p^2) \;,
	\label{jju2}
	\eeq
	with
	\beq
	\hat{\Pi}_{hh}(p^2)&=&\frac{3g^2}{{8 (4 \pi )^2}}\int_0^1 dx \,  \Bigg\{2  \xi  \left(m_h^2+p^2\right) \ln \left(\frac{m^2 \xi }{\mu ^2}\right)-2 \xi  m_h^2+2 \left(6 m^2-p^2\right) \ln \left(\frac{m^2}{\mu ^2}\right)\nonumber\\
	&-&(12 m^2-\frac{(m_h^4+2p^2 m_h^2)}{m^2}+4 p^2) \ln \left(\frac{m^2+p^2 (1-x) x}{\mu ^2}\right)-\frac{(2m_h^4+2p^2 m_h^2)}{m^2}\ln \left(\frac{m^2 \xi +p^2 (1-x) x}{\mu ^2}\right)\nonumber\\
	&-&12 m^2-2 \xi  p^2+2 p^2-\frac{ m_h^4}{m^2} \Big(-2 \ln \left(\frac{m_h^2}{\mu ^2}\right)+3 \ln \left(\frac{m_h^2+p^2 (1-x) x}{\mu ^2}\right)+2\Big)\Bigg\}
	\label{pihh}
	\eeq
	and
	\beq
	C_{hh}(p^2)= -\frac{3g^2}{{8 m^2(4 \pi )^2}}\int_0^1 dx \left( \ln \frac{p^2x(1-x)+m^2}{\mu^2} -\ln \frac{p^2x(1-x)+\xi m^2}{\mu^2} \right)  \;.
	\eeq
	Thus, for the one-loop Higgs propagator, we get
	\beq
	\braket{h(p)h(-p)}&=&\frac{1}{p^2+m_h^2- \hat{\Pi}_{hh}(p^2)} + C_{hh}(p^2) + \mathcal{O}(\hbar^2) \;.
	\label{hhresum}
	\eeq
	For the gauge  field, we split the
	two-point function into transverse and longitudinal parts in the usual way
	\beq
	\braket{A_{\mu}^a(p)A^b_{\nu}(-p)}= \braket{A_{\mu}^a(p)A^b_{\nu}(-p)}^T \mathcal{P}_{\mu \nu}(p)+ \braket{A_{\mu}^a(p)A^b_{\nu}(-p)}^L \mathcal{L}_{\mu \nu}(p),
	\eeq
	where we have introduced the transverse and longitudinal projectors, given respectively by
	\beq
	\mathcal{P}_{\mu\nu}(p)&=& \delta_{\mu\nu}-\frac{p_{\mu}p_{\nu}}{p^2}\,, \qquad	\mathcal{L}_{\mu\nu}(p)~=~ \frac{p_{\mu}p_{\nu}}{p^2}\,.
	\eeq
	We find
	\beq
	\braket{A_{\mu}^a(p)A_{\nu}^b(-p)}^T=\frac{\delta^{ab}}{p^2+m^2}+ \frac{\delta^{ab}}{(p^2+m^2)^2} \Pi_{AA^T}(p^2)+ \mathcal{O}(\hbar^2),
	\label{piAA}
	\eeq
	with $\Pi_{AA^T}(p^2)$ the one-loop correction to the self-energy calculated in Appendix \ref{Ap}. For $d=4-\epsilon$, following the procedure of dimensional regularization, we find that the divergent part for $\Pi_{AA^T}(p^2)$ is given by:
		\beq
		\Pi_{AA^T, \rm div}(p^2)&=&\frac{g^2}{\pi ^2 \epsilon } \left(-\frac{9 m^4}{16 m_h^2}-\frac{3 m_h^2}{32}-\frac{m^2 \xi }{8}-\frac{3 m^2}{32}-\frac{\xi  p^2}{8}+\frac{25 p^2}{48}\right),
		\eeq
		and these terms can be, following the  $\MSbar$-scheme, absorbed by means of appropriate counterterms. We remain with the finite part of the self-energy\footnote{For notational simplicity, we will call this finite part $\Pi_{AA^T}$ again. We will follow this notational convention as well for later correlation functions that we will encounter. }
	\beq
	\Pi_{AA^T}(p^2)&=& -\frac{ \delta^{ab} g^2 }{36 (4 \pi )^2 m^4 p^2 m_h^2}\int_0^1 dx \Bigg\{-27 m^4 p^2 m_h^4 \ln \left(\frac{m_h^2}{\mu ^2}\right)-27 m^6 \xi  p^2 m_h^2 \ln \left(\frac{m^2 \xi }{\mu ^2}\right)\nonumber\\
	&+&3 m^4 m_h^4 \left(m_h^2-m^2+2 p^2\right) \ln \left(\frac{m_h^2}{\mu ^2}\right)+27 m^4 p^2 m_h^2 \left(m_h^2+m^2 \xi \right)\nonumber\\
	&-&3 m^4 \xi  m_h^2 \left(2 m^4 (\xi -1)+m^2 (4 \xi +7) p^2+2 (\xi +9) p^4\right) \ln \left(\frac{m^2 \xi }{\mu ^2}\right)\nonumber\\
	&+&3 m^4 \ln \left(\frac{m^2}{\mu ^2}\right) \left(-m^2 m_h^4+m_h^2 \left(m^4 (2 \xi -1)+m^2 (4 \xi +45) p^2+2 (\xi +9) p^4\right)-54 m^4 p^2\right)\nonumber\\
	&+&m^4 \left(6 m^2 m_h^4+m_h^2 \left(3 m^4 (2 (\xi -2) \xi +1)+3 m^2 (\xi -1) (4 \xi -1) p^2+2 (3 \xi  (\xi +4)-17) p^4\right)-3 m_h^6+54 m^4 p^2\right)\nonumber\\
	&-&3 m_h^2 \Big[m^4 \left(2 p^2 \left(m_h^2-5 m^2\right)+\left(m_h^2-m^2\right){}^2+p^4\right) \ln \left(\frac{p^2 (1-x) x+m_h^2(1-x) +m^2 x}{\mu ^2}\right)\nonumber\\
	&-&2 \left(m^2+p^2\right)^2 \left(m^4 (\xi -1)^2+2 m^2 (\xi -5) p^2+p^4\right) \ln \left(\frac{p^2 (1-x) x+ \xi m^2 (1-x)+m^2 x}{\mu ^2}\right)\nonumber\\
	&+&p^2 \left(p^4-m^4\right) \left(4 m^2 \xi +p^2\right) \ln \left(\frac{p^2 (1-x) x+ \xi m^2}{\mu ^2}\right)\nonumber\\
	&+&p^2 \left(4 m^2+p^2\right) \left(12 m^4-20 m^2 p^2+p^4\right) \ln \left(\frac{p^2 (1-x) x+m^2}{\mu ^2}\right)\Big]\Bigg\}.
	\label{AAf}
	\eeq
	We see that \eqref{piAA} contains again terms of the order $ \frac{p^4}{(p^2+m^2)^2}\ln \left(\frac{m^2+p^2 (1-x) x}{\mu ^2}\right)$ and $ \frac{p^6}{(p^2+m^2)^2}\ln \left(\frac{m^2+p^2 (1-x) x}{\mu ^2}\right)$, which cannot be resummed for big values of $p^2$. We use
	\beq
	\frac{p^4}{(p^2+m^2)^2} \ln \frac{p^2x(1-x)+m^2}{\mu^2} &=& \ln \frac{p^2x(1-x)+m^2}{\mu^2} -\underline{\frac{(m^4+2p^2m^2)}{(p^2+m^2)^2} \ln \frac{p^2x(1-x)+m^2}{\mu^2}}
	\label{u1}
	\eeq
	and
	\beq
	\frac{p^6}{(p^2+m^2)^2} \ln \frac{p^2x(1-x)+m^2}{\mu^2} &=& (p^2-2m^2)\ln \frac{p^2x(1-x)+m^2}{\mu^2} +\underline{\frac{2m^6+3p^2m^4}{(p^2+m^2)^2} \ln \frac{p^2x(1-x)+m^2}{\mu^2}}
	\label{u2}.
	\eeq
	The underlined terms in \eqref{u1} and \eqref{u2} can be safely resummed. We rewrite
	\beq
	\frac{{\Pi}_{AA^T}(p^2)}{(p^2+m^2)^2}&=& \frac{\hat{\Pi}_{AA^T}(p^2)}{(p^2+m_h^2)^2}+C_{AA^T}(p^2),
	\label{jju2}
	\eeq
	with \allowdisplaybreaks
	\beq
	\hat{\Pi}_{AA^T}(p^2)&=& -\frac{ \delta^{ab} g^2 }{36 (4 \pi )^2 m^4 p^2 m_h^2}\int_0^1 dx \Bigg\{-27 m^4 p^2 m_h^4 \ln \left(\frac{m_h^2}{\mu ^2}\right)-27 m^6 \xi  p^2 m_h^2 \ln \left(\frac{m^2 \xi }{\mu ^2}\right)\nonumber\\
	&+&3 m^4 m_h^4 \left(m_h^2-m^2+2 p^2\right) \ln \left(\frac{m_h^2}{\mu ^2}\right)+27 m^4 p^2 m_h^2 \left(m_h^2+m^2 \xi \right)\nonumber\\
	&-&3 m^4 \xi  m_h^2 \left(2 m^4 (\xi -1)+m^2 (4 \xi +7) p^2+2 (\xi +9) p^4\right) \ln \left(\frac{m^2 \xi }{\mu ^2}\right)\nonumber\\
	&+&3 m^4  \left(-m^2 m_h^4+m_h^2 \left(m^4 (2 \xi -1)+m^2 (4 \xi +45) p^2+2 (\xi +9) p^4\right)-54 m^4 p^2\right)\ln \left(\frac{m^2}{\mu ^2}\right)\nonumber\\
	&+&m^4 \left(6 m^2 m_h^4+m_h^2 \left(3 m^4 (2 (\xi -2) \xi +1)+3 m^2 (\xi -1) (4 \xi -1) p^2+2 (3 \xi  (\xi +4)-17) p^4\right)-3 m_h^6+54 m^4 p^2\right)\nonumber\\
	&-&3 m_h^2 \Big[m^4 \left(2 p^2 \left(m_h^2-5 m^2\right)+\left(m_h^2-m^2\right){}^2+p^4\right) \ln \left(\frac{p^2 (1-x) x+(1-x) m_h^2+m^2 x}{\mu ^2}\right)\nonumber\\
	&-&2 m^4 (\xi -1)^2 \left(m^2+p^2\right)^2 \ln \left(\frac{p^2 (1-x) x+\xi  m^2 (1-x)+m^2 x}{\mu ^2}\right)\nonumber\\
	&+&\left(-2 m^4 (4 \xi -1) p^2 \left(m^2+p^2\right)\right) \ln \left(\frac{p^2 (1-x) x+\xi m^2}{\mu ^2}\right)\nonumber\\
	&+&(66 m^6 p^2-33 m^4 p^4) \ln \left(\frac{p^2 (1-x) x+m^2}{\mu ^2}\right)\Big]\Bigg\}.
	\label{hkg}
	\eeq
	and
	\beq
	C_{AA^T}(p^2)&=& \frac{ \delta^{ab} g^2 }{12 (4 \pi )^2 m^4  }\int_0^1 dx \Bigg\{(-4 m^2 (\xi -5)-2p^2) \nonumber\\
	&\times&\ln \left(\frac{p^2 (1-x) x+\xi m^2  (1-x)+m^2 x}{\mu ^2}\right)  \nonumber\\
	&+&\left(4 \,\xi m^2   +p^2-2m^2\right) \ln \left(\frac{p^2 (1-x) x+\xi m^2}{\mu ^2}\right)\nonumber\\
	&+&(-18 m^2 +p^2) \ln \left(\frac{p^2 (1-x) x+m^2}{\mu ^2}\right)\Bigg\}.
	\eeq
	Finally
	\beq
	\braket{A_{\mu}^a(p)A_{\nu}^b(-p)}^T=\delta^{ab}\left(  \frac{1}{p^2+m^2- \hat{\Pi}_{AA^T}(p^2)}+ C_{AA^T}(p^2)\right)  + \mathcal{O}(\hbar^2) \;.
	\label{piAAll}
	\eeq

	\section{One-loop evaluation of the correlation function of the local BRST invariant  composite operators  \label{IIIIr}}
	
	\subsection{Correlation function of the scalar BRST invariant composite operator $O(x)$}
	The gauge and BRST invariant local scalar composite operator $O(x)$ is given by
	\begin{equation}
	O(x)= \Phi^{\dagger}\Phi -\frac{v^2}{2} \;,\qquad s\; O(x) =0 \;,  \label{OOO}
	\end{equation}
	which, after using the expansion \eqref{expansion2}, becomes
	\beq
	O(x)&=& \frac{1}{2} \Big[\begin{pmatrix}
		1 & 0
	\end{pmatrix}((v+h)\textbf{1}-i\rho^a \tau^a))((v+h)\textbf{1}+i\rho^b \tau^b))\begin{pmatrix}
		1\\
		0
	\end{pmatrix}\Big]-\frac{v^2}{2}\nonumber\\
	&=&\frac{1}{2}\Big(h^2(x)+2 v h(x) + \rho^a (x) \rho^a(x)\Big) \;. \label{scopt}
	\eeq
	Notice that the operator $O(x)$ does not depend on the FP ghost field. Therefore,
	\beq
	\braket{{O}(x) {O}(y)} &=& v ^2 \braket{h(x) h(y)}+ v  \braket{h(x) \rho^b (y)\rho^b (y) }+ v  \braket{h(x) h(y)^2}+\frac{1}{4}\braket{h(x)^2 \rho^b (y)\rho^b (y)}+\frac{1}{4}\braket{h(x)^2 h(y)^2} \nonumber \\
	&+&\frac{1}{4}\braket{\rho^a (x)\rho^a (x) \rho^b (y)\rho^b (y)}. \label{exp1}
	\eeq
	Looking at the tree level expression of eq.~\eqref{exp1}, one easily obtains
	\beq
	\braket{{O}(p) {O}(-p)}_{\rm tree} = v ^2 \braket{h(p) h(-p)}_{\rm tree}
	= v^2 \frac{1}{p^2+m_h^2} \;, \label{treeoo}
	\eeq
	showing that the BRST invariant scalar operator $O(x)$ is directly linked to the Higgs propagator.
	
	Concerning now the one-loop calculation of expression \eqref{exp1}, after evaluating each term, see Appendix \ref{OO} for details, we find that the two-point correlation function of the scalar composite operator $O(x)$ develops a geometric series in the same way as the elementary field $h(x)$. This allows us the make a resummed approximation. Using  dimensional regularization in the  $\MSbar$-scheme, we find that, in the $R_{\xi}$-gauge
	\beq
	\braket{O(p)O(-p)}(p^2)&=& \frac{v^2}{p^2+m_h^2}+\frac{v^2 \,}{(p^2+m_h^2)^2}\Pi_{OO}(p^2)+\mathcal{O}(\hbar^2),
	\label{ark}
	\eeq
	with $\Pi_{OO}(p^2)$ the one-loop correction calculated in Appendix \ref{OO}. Following the procedure of dimensional regularization for $d=4-\epsilon$, we find that the divergent part of the one-loop correction is given by
	\beq
	\Pi_{OO, \rm div}^{-1}&=&\frac{1}{4 v^2 \pi ^2 \epsilon }\Big(\frac{9 g^4 p^2 v^2}{16 \lambda }+\frac{9 g^4 v^4}{16}+\frac{9}{8} g^2 p^2 v^2+p^4+\frac{1}{2} \lambda  p^2 v^2+\lambda ^2 v^4\Big),
	\label{el}
	\eeq
	which can be accounted for by appropriate counterterms, following the  $\MSbar$-scheme renormalization procedure. A full renormalization analysis will be presented elsewhere. Notice that expression \eqref{el}, when multiplied again with the factored-out $v^2$ of eq.~\eqref{ark},  is a polynomial in $p^2, v^2$, in full accordance with power counting renormalizability.
\\
	We remain with the finite part, given by
	
	\beq
	\Pi_{OO}(p^2)&=&\frac{1}{{32 v^2\pi ^2  m_h^2}}\int_0^1 dx \Bigg\{-24 m_h^2 m^4-6 m^2 p^2 \left(m_h^2+6 m^2\right) \ln \left(\frac{m^2}{\mu ^2}\right)\nonumber\\
	&-&m_h^2\left(p^2-2 m_h^2\right)^2 \ln \left(\frac{m_h^2 +p^2 x(1-x)}{\mu ^2}\right)\nonumber\\
	&-&3 m_h^2\left(12 m^4+4 m^2 p^2+p^4\right) \ln \left(\frac{m^2 +p^2 x(1-x)}{\mu ^2}\right)\nonumber\\
	&+&6 p^2 \left(m_h^4+m_h^2 m^2+2 m^4\right)-6 m_h^4 p^2 \ln \left(\frac{m_h^2}{\mu ^2}\right)\Bigg\}.
	\label{WW}
	\eeq
	Since \eqref{ark} contains terms of the order of $\frac{p^4}{(p^2+m^2)^2}\ln (p^2)$,
	we follow the steps \eqref{u1}-\eqref{jju2} to find the resummed correlation function in the one-loop approximation
	\beq
	G_{{O}{O}}(p^2)
	&=&\frac{v^2}{p^2+m_h^2-\hat{\Pi}_{OO}(p^2)}+C_{OO}(p^2)
	\label{dk3a}
	\eeq
	with
	\beq
	\hat{\Pi}_{OO}(p^2)&=&\frac{1}{{32 v^2\pi ^2  m_h^2}}\int_0^1 dx \Bigg\{-24 m_h^2 m^4-6 m^2 p^2 \left(m_h^2+6 m^2\right) \ln \left(\frac{m^2}{\mu ^2}\right)\nonumber\\
	&-&m_h^2(3 m_h^4 -6 m_h^2p^2) \ln \left(\frac{m_h^2 +p^2 x(1-x)}{\mu ^2}\right)\nonumber\\
	&-&3 m_h^2\left(12 m^4+4 m^2 p^2-m_h^4-2p^2m_h^2 \right) \ln \left(\frac{m^2 +p^2 x(1-x)}{\mu ^2}\right)\nonumber\\
	&+&6 p^2 \left(m_h^4+m_h^2 m^2+2 m^4\right)-6 m_h^4 p^2 \ln \left(\frac{m_h^2}{\mu ^2}\right)\Bigg\}
	\label{dk3b}
	\eeq
	and
	\beq
	C_{OO}(p^2)&=&-\frac{1}{32 \pi ^2  }
	\int_0^1 dx \Bigg\{\ln \left(\frac{m_h^2+p^2x(1-x)}{\mu^2}\right)+3\ln \left( \frac{ m^2 + p^2 x (1-x) }{\mu^2}\right)\Bigg\}.
	\label{grgra}
	\eeq
	Expressions \eqref{dk3a} and \eqref{grgra} show that, as expected, and unlike the Higgs propagator, eq.~\eqref{hhresum}, the correlator $\braket{O(p)O(-p)}$ is  independent from the gauge parameter $\xi$.

	\subsection{Vectorial composite operators}
	We identify three gauge invariant vector composite operators, following the definitions of 't Hooft in \cite{tHooft:1980xss}, namely
	\beq
	O_{\mu}^{3}&=&i\phi^{\dagger}D_{\mu}\phi,\nonumber\\
	O_{\mu}^{+}&=&\phi^{T}\begin{pmatrix}0 & 1\nonumber\\
		-1 & 0
	\end{pmatrix}D_{\mu}\phi,\\
	O_{\mu}^{-}&=&(O_{\mu}^{+})^{\dagger}.
	\eeq
	The gauge invariance of $O_{\mu}^{3}$ is apparent. For $O^{+}_{\mu}$, we can show the gauge invariance by using  the following $2 \times 2$ matrix representation of a generic $SU(2)$ transformation,
	\beq
	U=\begin{pmatrix}a & -b^{\star}\\
		b & a^{\star}
	\end{pmatrix}
	\eeq
	with determinant $\vert a \vert^2 +\vert b \vert^2=1$. Thus, we find that under a $SU(2)$ transformation
	\beq
	O_{\mu}^{+} &\rightarrow& (U\phi)^T \begin{pmatrix}0 & 1\\
		-1 & 0
	\end{pmatrix} D_{\mu}(U\phi)\nonumber\\
	&=& \phi^T U^T \begin{pmatrix}0 & 1\\
		-1 & 0
	\end{pmatrix} U D_{\mu}\phi\nonumber\\
	&=& \phi^T  \begin{pmatrix}0 & 1\\
		-1 & 0
	\end{pmatrix} D_{\mu}\phi=O_{\mu}^{+},
	\eeq
	which shows the gauge invariance of $O_{\mu}^{+}$ and, subsequently, of $O_{\mu}^{-}$. After using the expansion \eqref{expansion2}, the first composite operator reads
	\beq
	O_{\mu}^{3}&=&i \phi^{\dagger}D_{\mu}\phi
	\nonumber\\
	&=&i \phi^{\dagger}\partial_{\mu}\phi+\frac{1}{2}g\phi^{\dagger}\tau^{a}A_{\mu}^{a}\phi\nonumber\\
	&=&\frac{i}{2} \Bigg[(v+h)\partial_{\mu}h+i(v+h)\partial_{\mu}\rho^{3}-i\rho^{3}\partial_{\mu}h+\rho^{a}\partial_{\mu}\rho^{a}+i \rho^1 \partial_{\mu}\rho^2-i \rho^2 \partial_{\mu}\rho^1
	-\frac{i}{2}g(v+h)^{2}A_{\mu}^{3}+ig(v+h)(A_{\mu}^{1}\rho^{2}-A_{\mu}^{2}\rho^{1})\nonumber
	\\
	&+&\frac{i}{2}g\rho^{a}A_{\mu}^{3}\rho^{a}-ig\rho^{3}A_{\mu}^{b}\rho^{b}\Bigg]\nonumber\\
	&=& \frac{1}{2} \Bigg[-(v+h)\partial_{\mu}\rho^{3}+\rho^{3}\partial_{\mu}h- \rho^1 \partial_{\mu}\rho^2+ \rho^2 \partial_{\mu}\rho^1
	+\frac{1}{2}g(v+h)^{2}A_{\mu}^{3}-g(v+h)(A_{\mu}^{1}\rho^{2}-A_{\mu}^{2}\rho^{1})\nonumber
	\\
	&-&\frac{1}{2}g\rho^{a}A_{\mu}^{3}\rho^{a}+g\rho^{3}A_{\mu}^{b}\rho^{b}\Bigg]+ \frac{i}{2} \partial_{\mu}O,
	\eeq
	and since the last term, {\it i.e.}~$\frac{i}{2} \partial_{\mu}O$, is  BRST invariant, the sum of the others terms has to be  BRST invariant too. Therefore, we can introduce the following three  ``reduced'' vector composite operators $R^a_\mu$ with  $a=1,2,3$ :
	\beq
	R_{\mu}^{1}&=&\frac{i}{2}\Big( O_{\mu}^+ - O^-_{\mu}\Big),\nonumber\\
	R_{\mu}^{2}&=&\frac{1}{2}\Big( O_{\mu}^+ + O^-_{\mu}\Big),\nonumber\\
	R_{\mu}^{3}&=&O_{\mu}^3-\frac{i}{2}\partial_{\mu}O,  \label{threeop}
	\eeq
	so that
	\beq
	R_{\mu}^a &=& \frac{1}{2}\Bigg[-(v+h)\partial_{\mu}\rho^a+\rho^a \partial_{\mu} h - \varepsilon^{abc} \rho^b \partial_{\mu} \rho^c
	+\frac{1}{2}g (v+h)^2 A_{\mu}^a-g(v+h)\varepsilon^{abc}(\rho^b A_{\mu}^c)\nonumber\\
	&-& \frac{1}{2}g A_{\mu}^a \rho^m \rho^m +g \rho^a A_{\mu}^m \rho^m\Bigg]\;,  \label{threeop1}
	\eeq
	with $R^a_{\mu}$ both gauge and BRST invariant, thus
	\begin{equation}
		s R^a_\mu(x) = 0 \;. \label{BrstRR}
	\end{equation}
Notice that, as in the case of the scalar operator $O(x)$, the quantities $R^a_{\mu}(x)$ do not contain FP ghosts. Remarkably, the BRST  invariant operators $R_{\mu}^a$ transform like a triplet under the custodial symmetry \eqref{custodial}, namely
	\beq
	\delta R_{\mu}^a= \varepsilon^{abc}\beta^b R_{\mu}^c \;, \label{tripr}
	\eeq	
In work in progress, we will confirm that the operators $R_\mu^a$ are the unique dimension 3 triplet vector operators belonging to the non-trivial BRST-cohomology and that they renormalize in the expected fashion \cite{Collins:1984xc}, that is, up to BRST-exact terms and terms proportional to the equations of motion, complemented with contact terms to render their correlations fully finite, thereby generalizing the construct of \cite{Capri:2020ppe}. We stress here that the explicit BRST cohomology construction in the non-Abelian vector sector is a tedious non-trivial exercise, hence subject of a separate paper. Needless to say, similar comments apply to the scalar operator $O$. 

Since the only rank two invariant tensor is $\delta^{ab}$, we can write, moving to momentum space,
	\beq
	\braket{ R^a_{\mu}(p)\,R^b_{\nu}(-p)}= \delta^{ab} R_{\mu \nu}(p^2 )\,\, \rightarrow \,\, R_{\mu \nu}(p^2 )= \frac{1}{3} \braket{ R^a_{\mu}(p)\,R^a_{\nu}(-p)}, \label{cc1}
	\eeq
	as well as
	\beq
	R_{\mu \nu}(p^2)= R(p^2) \mathcal{P}_{\mu \nu} (p) + L (p^2) \mathcal{L}_{\mu \nu}(p), \label{cc2}
	\eeq
	so that  in $d$ dimensions,
	\beq
	R(p^2 )=\frac{1}{3}\frac{\mathcal{P}_{\mu \nu}(p)}{(d-1)}\braket{ R^a_{\mu}(p)\,R^a_{\nu}(-p)}, \label{cc3}
	\eeq
	and
	\beq
	L(p^2)= \frac{1}{3} \mathcal{L}_{\mu \nu}(p)\braket{ R^a_{\mu}(p)\,R^a_{\nu}(-p)}. \label{cc4}
	\eeq
	One recognizes that eqs.\eqref{cc1}-\eqref{cc4} display exactly the same structure of the gauge vector boson correlation function $\langle A^a_\mu(p) A^a_\nu(-p)\rangle$.
	
	In the $R_{\xi}$-gauge, the non-vanishing contributions, up to first order in $\hbar$, to the correlation function $\braket{ R^a_{\mu}(p)\,R^b_{\nu}(-p)}$ are
	\beq
	\braket{R^a_{\mu}(p)\,R^a_{\nu}(-p)}&=&\frac{1}{16}g^{2}v^{4}\langle A_{\mu}^{a}(p)\, A_{\nu}^{a}(-p)\rangle-\langle(\rho^{a}\partial_{\mu}h)(p)\, (\partial_{\nu}\rho^{a}h)(-p)\rangle+\frac{1}{4}p_{\mu}p_{\nu}\langle(\rho^{a}h)(p) \,(\rho^{a}h)(-p)\rangle \nonumber\\
	&+&\frac{1}{8}p_{\mu}p_{\nu}\langle(\rho^{a}\rho^{b})(p)\, (\rho^{a} \rho^{b})(-p)\rangle-\frac{1}{2}\langle(\rho^{a}\partial_{\mu}\rho^{b})(p)\,(\partial_{\nu}\rho^{a}\rho^{b})(-p)\rangle\nonumber\\
	&+&\frac{1}{4}g^{2}v^{3}\langle A_{\mu}^{a}(p)\, (A_{\nu}^{a}h)(-p)\rangle-\frac{i}{4}gv^{3}p_{\nu}\langle A_{\mu}^{a}(p)\, \rho^{a}(-p)\rangle
	\nonumber\\
	&+&\frac{1}{4}v^{2}p_{\mu}p_{\nu}\langle\rho^{a}(p) \rho^{a}(-p)\rangle+\frac{1}{6}g^{2}v^{2}\langle(\rho^{a}A_{\mu}^{b})(p)\, (\rho^{a}A_{\nu}^{b})(-p)\rangle\nonumber\\
	&-&\frac{1}{24}g^{2}v^{2}\langle(\rho^{a}\rho^{a}A_{\mu}^{b})(p)\, A^{b}(-p)\rangle+\frac{1}{8}g^{2}v^{2}\langle (h^2 A_{\mu}^{a})(p) \,A_{\nu}^{a}(-p)\rangle
	\nonumber\\
	&+&\frac{1}{4}g^{2}v^{2}\langle (h A_{\mu}^{a})(p) \,(h A_{\nu}^{a})(-p)\rangle+\frac{i}{4}v^{2}gp_{\mu}\langle (h\rho^{a})(p)\,A_{\nu}^{a}(-p)\rangle
	\nonumber\\
	&+&\frac{1}{2}gv^{2}\langle(\partial_{\mu}h\rho^{a})(p)\,A_{\nu}^{a}(-p)\rangle+\frac{i}{2}gv^{2}p_{\mu}\langle\rho^{a}(p)\,(h A_{\nu}^{a})(-p)\rangle\nonumber\\
	&-&\frac{1}{4}gv^{2}\varepsilon^{abc}\langle A^{a}(p)\,(\rho^{b}\partial_{\mu}\rho^{c})(-p)\rangle-\frac{ig v^{2}}{2}\varepsilon^{abc}p_{\mu}\langle\rho^{a}(p)\,(\rho^{b} A_{\nu}^{c})(-p)\rangle\nonumber
	\\
	&+&\frac{1}{2}vp_{\mu}p_{\nu}\langle (h\rho^{a})(p)\,\rho^{a}(-p)\rangle-ivp_{\nu}\langle(\partial_{\mu}h\rho^{a})(p)\,\rho^{a}(-p)\rangle
	\label{OO+}
	\eeq
	The first term in expression \eqref{OO+} is the gauge field propagator $\braket{A_{\mu}^a (p)A_{\mu}^a(-p)} $, which means that $R^a_{\mu}$ can be thought as a kind of  BRST invariant extension of the elementary gauge field $A_{\mu}^a$. At tree-level, we find in fact
	\beq
\braket{R^a_{\mu}(p)\,R^a_{\nu}(-p)}_{tree}&=& \frac{1}{16}g^{2}v^{4}\langle A_{\mu}^{a}(p) A_{\nu}^{a}(-p)\rangle+\frac{1}{4}v^{2}\partial_{\mu}\partial_{\nu}\langle\rho^{a}(p) \rho^{a}(-p)\rangle\nonumber\\
	&=& \frac{3}{16}g^2 v^4 \frac{1}{p^2+m^2}\mathcal{P}_{\mu\nu}(p)+\frac{3}{4}v^2 \mathcal{L}_{\mu\nu}(p),
	\eeq
	where we can see that, apart from the constant factor $\frac{3}{4}v^2$ appearing in the longitudinal sector, the transverse component reproduces exactly the transverse gauge tree-level propagator.
	
Collecting the results from Appendix~\ref{RRR}, we see that the transverse part of the correlator \eqref{cc3} in the $R_{\xi}$-gauge is
	\beq\label{arkbis}
	R(p^2)&=& \frac{1}{16}g^2 v^4\Bigg(\frac{1}{p^2+m^2}+\frac{1}{(p^2+m^2)^2} \Pi_R(p^2) \Bigg)+ \mathcal{O}(\hbar^2)
	\eeq
	where the divergent part of the one-loop correction is
	\beq
	\Pi_{R,\rm div}(p^2)&=&\frac{g^2}{\epsilon \pi^2}\left(- \frac{m_h^2p^4}{32 m^4}+\frac{9 m^4}{16m_h^2}+\frac{9 m^2 p^2}{8 m_h^2}+\frac{m_h^2p^2}{8 m^2}+\frac{m_h^2}{16}-\frac{p^6}{48 m^4}+\frac{23p^4}{96 m^2}+\frac{7p^2}{8}\right),
	\label{el4}
	\eeq
which is accounted for by appropriate counterterms in the $\MSbar$-scheme renormalization procedure. Notice that, compatible with power counting, expression \eqref{el4} is again a polynomial in $p^2, v^2$ upon re-instating the factored-out $v^4$ of eq.~\eqref{arkbis}.
\\
We remain with
	\beq
	\Pi_{R}(p^2)&=&\frac{3}{36 \pi^2 g^2 v^4 m_h^2}\int_0^1 dx \Bigg\{ 6 m^4 \left(m_h^4+3 m^4\right)-\frac{p^4 m_h^2}{3} \left(9 m_h^2+35 m^2+4 p^2\right)\nonumber\\
	&+& p^2 \left(m^4 m_h^2+10 m^2 m_h^4+m_h^6+36 m^6\right)+m_h^4 \left(-p^2 \left(m_h^2+11 m^2\right)-6 m^4+p^4\right) \ln \left(\frac{m_h^2}{\mu ^2}\right)\nonumber\\
	&+&m_h^2 \left(2 p^4 \left(m_h^2-5 m^2\right)+\left(m^2-m_h^2\right)^2 p^2+p^6\right) \ln \left(\frac{p^2(1-x)x+(1-x)m_h^2+x m^2}{\mu ^2}\right)\nonumber\\
	&+&m_h^2\left(48 m^6-68 m^4 p^2-16 m^2 p^4+p^6\right) \ln \left(\frac{p^2 (1-x) x+m^2}{\mu ^2}\right)\nonumber\\
	&+& m^2 \left(m_h^2 \left(-48 m^4-17 m^2 p^2+3 p^4\right)+p^2 m_h^4-54 \left(m^6+2 m^4 p^2\right)\right) \ln \left(\frac{m^2}{\mu ^2}\right)\Bigg\}.
	\label{GG}
	\eeq

Since \eqref{GG} contains terms of the order of $\frac{p^4}{(p^2+m^2)^2}\ln (p^2)$ and $\frac{p^6}{(p^2+m^2)^2}\ln (p^2)$,
	we follow the steps \eqref{u1}-\eqref{jju2} to find the resummed   propagator in the one-loop approximation, namely
	\beq
	G_{R}(p^2)
	&=&\frac{1}{16}g^2 v^4\Big(\frac{1}{p^2+m_h^2-\hat{\Pi}_{R}(p^2)}\Big)+C_{R}(p^2)
	\label{dk3}
	\eeq
	with
	\beq
	\hat{\Pi}_{R}(p^2)&=&\frac{3}{36 \pi^2 g^2 v^4 m_h^2}\int_0^1 dx \Bigg\{ 6 m^4 \left(m_h^4+3 m^4\right)-\frac{p^4 m_h^2}{3} \left(9 m_h^2+35 m^2+4p^2\right)\nonumber\\
	&+& p^2 \left(m^4 m_h^2+10 m^2 m_h^4+m_h^6+36 m^6\right)+m_h^4 \left(-p^2 \left(m_h^2+11 m^2\right)-6 m^4+p^4\right) \ln \left(\frac{m_h^2}{\mu ^2}\right)\nonumber\\
	&+& m_h^2\left(m_h^4 p^2-2 m_h^2 m^4-6 m_h^2 m^2 p^2+12 m^6+24 m^4 p^2\right) \ln \left(\frac{x \left(m^2-p^2 (x-1)\right)-(x-1) m_h^2}{\mu ^2}\right)\nonumber\\
	&+&33 m_h^2\left(2 m^6- m^4 p^2\right) \ln \left(\frac{m^2-p^2 (x-1) x}{\mu ^2}\right)\nonumber\\
	&+& m^2 \left(m_h^2 \left(-48 m^4-17 m^2 p^2+3 p^4\right)+p^2 m_h^4-54 \left(m^6+2 m^4 p^2\right)\right) \ln \left(\frac{m^2}{\mu ^2}\right)\Bigg\}
	\label{dk3u}
	\eeq
	and
	\beq
	C_{R}(p^2)&=&\frac{1}{ 12(4\pi) ^2   }
	\int_0^1 dx \Bigg\{\left(-18 m^2 +p^2\right) \ln \left(\frac{p^2 (1-x)x + m^2}{\mu ^2}\right)\nonumber\\
	&+& \Big((2 \left(m_h^2-6 m^2\right)+p^2) \ln \left(\frac{p^2x(1-x)+(1-x)m_h^2+x m^2}{\mu ^2}\right)\Bigg\}.
	\label{grgr}
	\eeq
	Looking now at the longitudinal part $L(p^2)$, it turns out to be
	\beq
	L(p^2)&=&\frac{1}{4} v^2	-\frac{1}{(4 \pi) ^2 }\Bigg(\frac{m_h^4-3m_h^4 \ln \left(\frac{m_h^2}{\mu ^2}\right)+9m^4-27 m^4 \ln \left(\frac{m^2}{\mu ^2}\right)}{2  m_h^2}  -\frac{1}{\epsilon} \left( m_h^2-9\frac{ m^4}{m_h^2}\right) \Bigg) \;.  \label{lexp}
	\eeq
	As it happens in the tree-level case, expression \eqref{lexp} is independent from the momentum $p^2$, meaning that it does not correspond to the propagation of some physical mode, a feature which is expected to be valid in general. This will be discussed elsewhere, as it requires an in-depth analysis of non-trivial Ward identities and their consequent restrictions on the correlation functions of the composite operators.

	\section{Spectral properties \label{V}}
	In this section, we will calculate the spectral properties associated with the correlation functions obtained in the last section. In   subsection \ref{2a}, we will shortly remind the techniques employed in  \cite{Dudal:2019aew} to obtain the pole mass, residue and spectral density up to first order in $\hbar$. In  subsection \ref{2b}, we analyze the spectral properties of the elementary fields. In \ref{compp}, the spectral properties of the composite operators $O(x)$ and $R^a_{\mu}(x)$ are discussed.

	\subsection{Obtaining the spectral function \label{2a}}
	We compare the K\"all\'{e}n-Lehmann \footnote{ We remind here that in the case of higher dimensional operators, the spectral representation,
		eq.~\eqref{ltt}, might require appropriate subtraction terms in order to ensure convergence. A standard way to cure this problem is to
		subtract from $G(p^2)$ the first few terms of its Taylor expansion at $p=0$ \cite{colangelo2001qcd}, leading to $ G^{(n)}(p^2)=(-p^2)^n \int dt \frac{\rho(t)}{(t^n(t+p^2))}$ in the $n$th subtraction, making the integral more and more convergent.
		In our theory we can make use of the subtracted equations at $p^2=0$ because all fields are massive in the $R_{\xi}$-gauge, so there are no divergences at zero momentum. These polynomial subtraction terms are in a direct correspondence with the aforementioned contact terms necessary to render the composite operator's correlation functions finite, see \cite{Capri:2020ppe} for more on this.} spectral representation for the propagator of a generic field $\widetilde{O}(p)$
	\beq
	\braket{\widetilde{O}(p)\widetilde{O}(-p)}=G(p^2)=\int_0^{\infty} dt \frac{\rho (t)}{t+p^2},
	\label{ltt}
	\eeq
	where $\rho(t)$ is the spectral density function and $G(p^2)$ stands for the  resummed propagator
	\beq
	G(p^2)&=& \frac{1}{p^2+m^2-\Pi(p^2)}.
	\label{iuh}
	\eeq
	The pole mass for any massless or massive field excitation is obtained by calculating the pole of the resummed propagator, that is, by solving
	\beq
	p^2+m^2-\Pi(p^2)=0 \, \label{ppp}
	\eeq
	and its solution defines the pole mass $p^2=-m_{\rm pole}^2$. As consistency requires us to work up to a fixed order in perturbation theory, we should solve eq.~\eqref{ppp} for the pole mass in an iterative fashion. Therefore, to first order in  $\hbar$ , we find
	\beq
	m_{\rm pole}^2=m^2-\Pi^{\rm 1-loop}(-m^2)+\mathcal{O}(\hbar^2),
	\label{pol}
	\eeq
	where $\Pi^{\rm 1-loop}$ is the first order, or one-loop, correction to the propagator.
	Now, we write eq.~\eqref{iuh} in a slightly different way, namely
	\beq
	G(p^2)&=& \frac{1}{p^2+m^2-\Pi(p^2)}\nonumber\\
	&=&\frac{1}{p^2+m^2-\Pi^{\rm 1-loop}(-m^2)-(\Pi(p^2)-\Pi^{\rm 1-loop}(-m^2))}\nonumber\\
	&=&\frac{1}{p^2+m_{\rm pole}^2-\widetilde{\Pi}(p^2)},
	\label{op}
	\eeq
	where we defined $\widetilde{\Pi}(p^2)=\Pi(p^2)-\Pi^{\rm 1-loop}(-m^2)$. At one-loop, expanding  $\widetilde{\Pi}(p^2)$ around $p^2=-m^2_{\rm pole}=-m^2+\mathcal{O}(\hbar)$  gives the residue
	\beq
	Z&=& \lim_{p^2 \rightarrow -m_{\rm pole}^2} (p^2+m_{\rm pole}^2)G(p^2) \nonumber \\
	&=&\frac{1}{1-\partial_{p^2} \Pi(p^2)\vert_{p^2=-m^2}} \nonumber \\
	&=&1+\partial_{p^2} \Pi(p^2)\vert_{p^2=-m^2}+\mathcal{O}(\hbar^2).
	\label{kaf}
	\eeq
	We now write \eqref{op} to first order in $\hbar$ as
	\beq
	G(p^2)&=&\frac{Z}{(p^2+m_{\rm pole}^2-\widetilde{\Pi}(p^2))Z}\nonumber\\
	&=&\frac{Z}{p^2+m_{\rm pole}^2-\widetilde{\Pi}(p^2)+(p^2+m_{\rm pole}^2)\frac{\partial \widetilde{\Pi}(p^2)}{\partial p^2}\vert_{p^2=-m^2}}\nonumber\\
	&=&\frac{Z}{p^2+m_{\rm pole}^2}+Z\left(\frac{\widetilde{\Pi}(p^2)-(p^2+m_{\rm pole}^2)\frac{\partial \widetilde{\Pi}(p^2)}{\partial p^2}\vert_{p^2=-m^2}}{(p^2+m_{\rm pole}^2)^2}\right),
	\label{12}
	\eeq
	where in the last line we used a first-order Taylor expansion so that the propagator has an isolated pole at $p^2=-m_{\rm pole}^2$. In  \eqref{ltt} we can isolate this pole in the same way, by defining the spectral density function as  $\rho(t)=Z \delta(t-m_{\rm pole}^2)+\widetilde{\rho}(t)$, giving
	\beq
	G(p^2)=\frac{Z}{p^2+m_{\rm pole}^2}+ \int_0^{\infty} dt\frac{\widetilde{\rho}(t)}{t+p^2}
	\label{22}
	\eeq
	and we identify the second term in each of the representations \eqref{12} and \eqref{22} as the $\textit{reduced propagator}$
	\beq
	\widetilde{G}(p^2)&\equiv& G(p^2)-\frac{Z}{p^2+m_{\rm pole}^2},
	\eeq
	so that
	\beq
	\widetilde{G}(p^2)= \int_0^{\infty}dt \frac{\widetilde{\rho}(t)}{t+p^2} &=&Z\left(\frac{\widetilde{\Pi}(p^2)-(p^2+m_{\rm pole}^2)\frac{\partial \widetilde{\Pi}(p^2)}{\partial p^2}\vert_{p^2=-m^2}}{(p^2+m_{\rm pole}^2)^2}\right).
	\label{pp}
	\eeq
	Finally, using Cauchy's integral theorem   from complex analysis, we can find the spectral density $\widetilde{\rho}(t)$ as a function of $\widetilde{G}(p^2)$, giving
	\beq
	\widetilde{\rho}(t)=\frac{1}{2\pi i}\lim_{\epsilon\to 0^+}\left(\widetilde{G}(-t-i\epsilon)-\widetilde{G}(-t+i\epsilon)\right).
	\label{key}
	\eeq
	
	\subsubsection{Pole masses}
	There is an interesting consequence  of the definition of the first-order pole mass, eq.~\eqref{pol}. When calculating one-loop corrections to the two-point function of the composite operators $O(p)$ and $R_{\mu}^a(p)$, we find that
	\beq
	\Pi_{\rm composite}(p^2)&=& \Pi_{\rm elementary}(p^2)+ \Pi_{\rm 1-leg}(p^2) (p^2+m^2)+\Pi_{\rm 0-leg}(p^2)(p^2+m^2)^2,
	\eeq
	where $\Pi_{\rm 1-leg}(p^2)$ and $\Pi_{\rm 0-leg}(p^2)$ are the composite one-loop contributions to the correction of the composite field's two-point functions, with one external leg and zero external legs, respectively. From this, we see immediately that
	\beq
	\Pi_{\rm composite}(-m^2)&=& \Pi_{\rm elementary}(-m^2)
	\eeq
	and therefore, up to first order in $\hbar$, we find
	\beq
	m^2_{\rm pole, composite}&=&m^2_{\rm pole, elementary}
	\label{zlf}
	\eeq
	which means that the elementary operators and their composite extensions share the same mass. This is an important feature, providing an alternative way to the Nielsen identities, to understand  why the pole masses of the elementary particles are gauge invariant  and not just gauge parameter independent.

	\subsection{Spectral properties of the elementary fields \label{IV}}
	\begin{table}[h!]
		\begin{tabular}{|c|c|c|}
			\hline
			& \textbf{Region I} & \textbf{Region II} \\ \hline
			$~v~$  & 0.8 $\mu$                     & 1 $\mu$   \\ \hline
			$~g~$  & 1.2                          & 0.5           \\ \hline
			$~\lambda~$  & 0.3                          & 0.205    \\ \hline
		\end{tabular}		\caption{Parameter values used in the spectral density functions.}
		\label{tabel}
	\end{table}
	We first discuss the spectral properties of the elementary fields: the scalar Higgs field $h(x)$ and  the transverse part of the gauge field $A^a_{\m}(x)$. We will work with two sets of parameters, set out in Table \ref{tabel}. All values are given in units of the energy scale $\mu$. Also, we have that $m^2= \frac{1}{4}g^2 v^2$ and $m_h^2=\lambda v^2$, so that $m^2=0.23\, \mu^2$ and $m_h^2=0.192 \,\mu^2$ in Region I and    $m^2=0.625\, \mu^2$ and $m_h^2=0.205  \,\mu^2$ in Region II.

	\subsubsection{The Higgs field}
	
\begin{figure}[H]
		\centering
		\includegraphics[width=10
		cm]{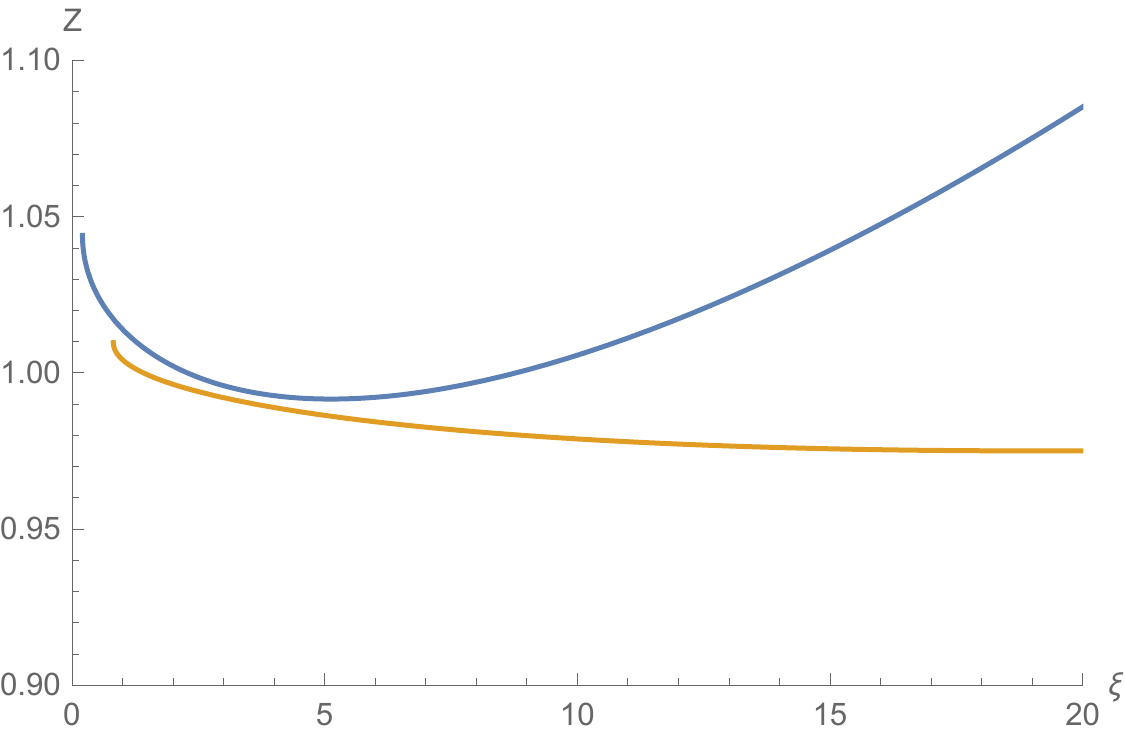}
		\caption{Dependence of the residue $Z$  for the Higgs field propagator on the gauge parameter $\xi$, for Region I (Blue), and Region II (Orange). }
		\label{ZZ}
	\end{figure}

	For the Higgs fields, following the steps from section \ref{2a}, we find the pole mass to first order in $\hbar$ to be: for Region I
	\beq
	m_{h,\rm pole}^2 &=& 0.207 \, \mu^2,
	\label{higgsmass}
	\eeq
	and for Region II
	\beq
	m_{h,\rm pole}^2 &=& 0.206 \, \mu^2, \label{higgsmass2}
	\eeq
	for all values of the parameter $\xi$. This means that while the Higgs propagator \eqref{hhf} is gauge dependent, the pole mass is gauge independent. This is in full agreement  with the Nielsen identities of the $SU(2)$ Higgs model studied in \cite{gambino1999fermion}. The residue, however, is gauge dependent, as is depicted in  FIG.~\ref{ZZ}.
	For small values of $\xi$, including the Landau gauge $\xi=0$, the residue is not well-defined, and we cannot determine the spectral density function, as we will explain further in the next section.
	
\begin{figure}[H]
		\centering
		\includegraphics[width=18cm]{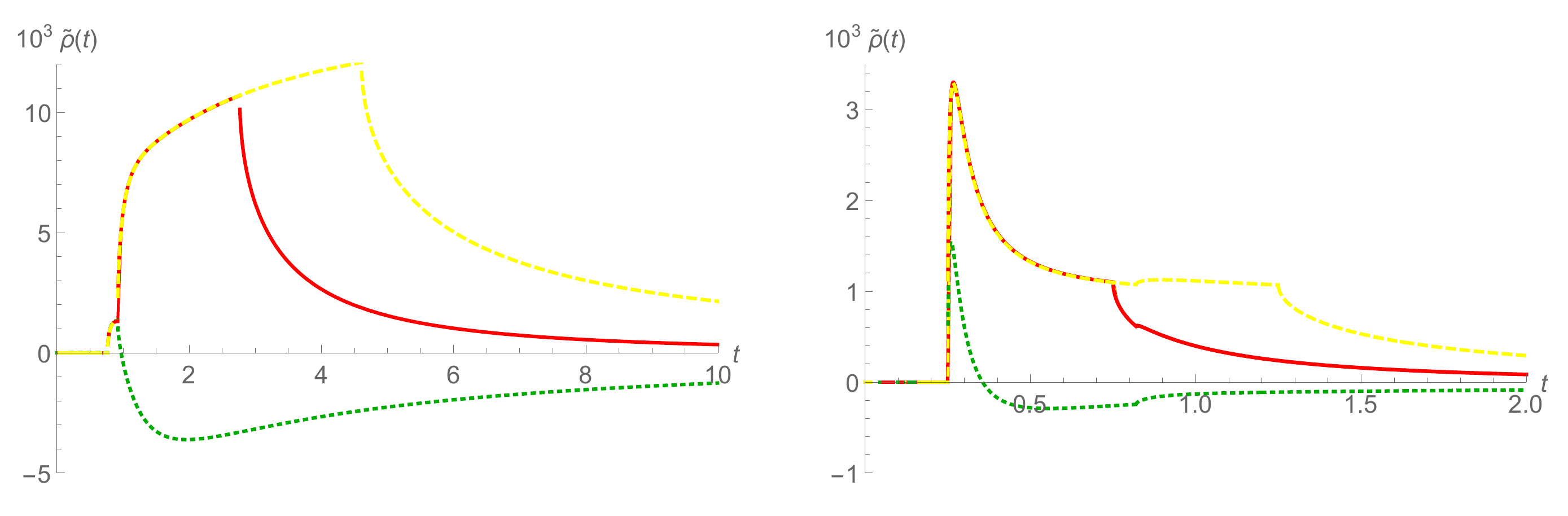}
		\caption{Spectral functions for the propagator $\langle h(p)h(-p) \rangle$, for $\xi = 1$ (Green, Dotted), $\xi = 3$ (Red, Solid),
			$\xi= 5$ (Yellow, Dashed),  with $t$ given in units of $\mu^2$, for Region I (left) and Region II (right) with the parameter values given in Table \ref{tabel}. }
		\label{Y33}
	\end{figure}

	In  FIG.~\ref{Y33}, we find the spectral density functions both regions, for three values of $\xi: 1, 2, 5$. Looking at Region I, we see the first two-particle state appearing at $t=(m_h+m_h)^2=0.768\,  \mu^2$, followed by another two-particle state at $t=(m+m)^2=0.922 \, \mu^2$. Then, we see that there is a negative contribution, different for each diagram, at $t=(\sqrt{\xi} m+\sqrt{\xi} m)^2$. This corresponds to the (unphysical) two-particle state of two Goldstone bosons. For $\xi < 3$, this leads to a negative contribution for the spectral function, probably due to the large-momentum behaviour of the Higgs propagator \eqref{hhf}, for a detailed discussion see \cite{Dudal:2019aew}. For Region II, we find essentially the same behaviour: a Higgs two-particle state at $t=(m_h+m_h)^2=0.81\,  \mu^2$, and a gauge field two-particle state at $t=(m+m)^2=0.25\, \mu^2$. We also see a negative contribution different for each diagram at $t=(\sqrt{\xi} m+\sqrt{\xi} m)^2$, corresponding to the (unphysical) two-particle state of two Goldstone bosons.

	\subsubsection{The transverse gauge field \label{2b}}
\begin{figure}[H]
		\centering
		\includegraphics[width=10
		cm]{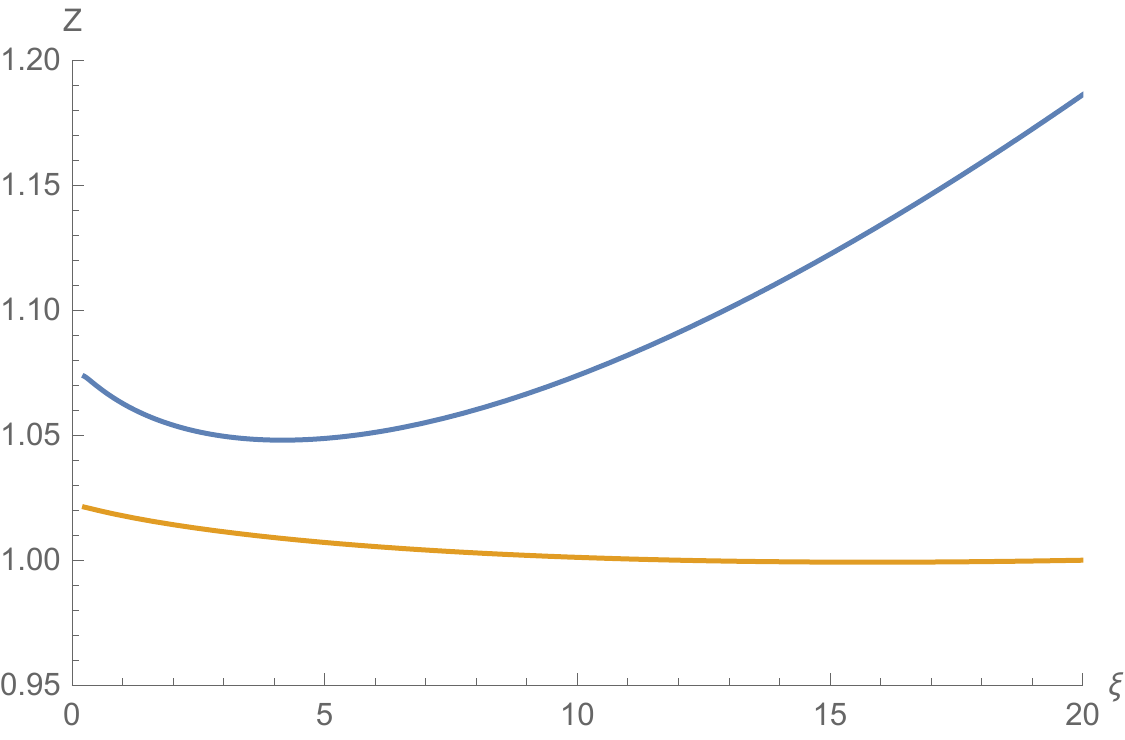}
		\caption{Dependence of the residue $Z$  for the gauge field propagator from  the gauge parameter $\xi$, for Region I (Blue), and Region II (Orange). }
		\label{Z}
	\end{figure}

	For the gauge field propagator, following the steps  from section~\ref{2a}, we find the first-order pole mass of the transverse gauge field to be: for Region I
	\beq
	m_{\rm pole}^2 &=& 0.274 \, \mu^2
	\label{polem}
	\eeq
	and for Region II
	\beq
	m_{\rm pole}^2 &=& 0.065 \, \mu^2
	\label{polema}
	\eeq
	for all values of the parameter $\xi$, so that the pole mass is independent from the gauge parameter. The residue is, however,  gauge dependent as is depicted in  FIG.~\ref{Z}. For small values of $\xi$ the residue is not well-defined, as we will explain further in the next section.
	
\begin{figure}[H]
		\centering
		\includegraphics[width=18cm]{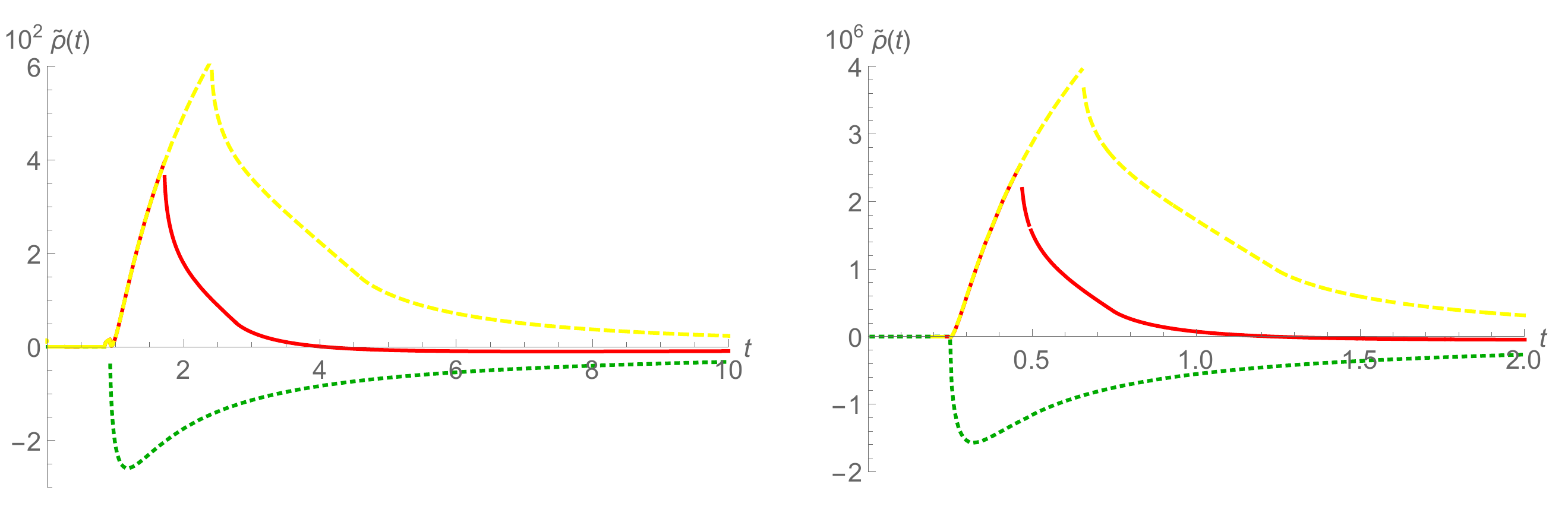}
		\caption{Spectral functions for the propagator $\langle A^a_{\mu}(p)A^b_{\nu}(-p) \rangle^T$, for $\xi = 1$ (Green, Dotted), $\xi = 3$ (Red, Solid),
			$\xi= 5$ (Yellow, Dashed),  with $t$ given in units of $\mu^2$, for Region I (left) and Region II (right) with the parameter values given in Table \ref{tabel}.}
		\label{Y3}
	\end{figure}

	In  FIG.~\ref{Y3}, we find the spectral density functions for both regions, for three values of $\xi$: $1, 2, 5$. Looking at Region I, we see the first two-particle state appearing at $t=(m_h+m)^2=0.843\,  \mu^2$, followed by a two-particle state at $t=(m+m)^2=0.922 \, \mu^2$. Then, we see that there is a negative contribution, different for each diagram, at $t=( m+\sqrt{\xi} m)^2$. This corresponds to the (unphysical) two-particle state of a  gauge and Goldstone boson. For Region II, we find a gauge field two-particle state at $t=(m+m)^2=0.25 \, \mu^2$. We also see a negative contribution different for each diagram at $t=( m+\sqrt{\xi} m)^2$, corresponding to the (unphysical) two-particle state of two Goldstone bosons.

	\subsection{Unphysical  threshold effects \label{stable}}
	
	\begin{figure}[H]
		\centering
		\includegraphics[width=8cm]{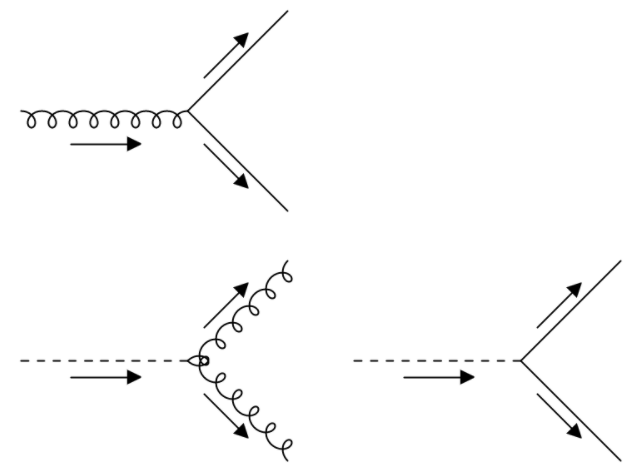}
		\caption{Possible decays of the  gauge boson (above) and the Higgs boson (below). The arrows indicate the momentum flow.}
		\label{yy}
	\end{figure}
	From the Feynman vertex rules given in Appendix~\ref{appA}, for certain values of the masses, unphysical  threshold effects can occur. These effects imply that for certain values of the (physical and unphysical) parameters, a ``decay'' occurs of a gauge and Higgs boson into two other particles, see  FIG.~\ref{yy}. We distinguish three cases:
	
	\begin{itemize}
		\item[$(1.)$] Decay of a gauge vector boson in two Goldstone bosons: this happens when $m > 2  \sqrt{\xi} m$.
		\item[$(2.)$] Decay of a Higgs boson in two gauge vector bosons: this happens when $m_h > 2 m$.
		\item[$(3.)$] Decay of a Higgs boson in two Goldstone bosons: this happens when $m_h > 2 \sqrt{\xi} m$.
	\end{itemize}
	In order to guarantee the stability of the gauge boson, we therefore need from $(1.)$ that $\xi > \frac{1}{4}$. This means that for the Landau gauge $\xi =0$, the elementary gauge boson is not stable.
	For the Higgs particle, to guarantee stability we need from $(2.)$ that $m_h < 2m$. Then, from $(3.)$ we find that $\xi >  \frac{m_h^2}{4m^2}$. This is the window in which we can work with a stable model.		We can have a look at what happens when we go outside of this window. For the Higgs particle, we see that for $m_h > 2m$, or $\lambda < g^2$, we will find a complex value for the first order pole mass, calculated through \eqref{ppp}. For $\lambda \geq g^2$, we will always find a real pole mass. Since the pole mass is gauge invariant, we find that this is true for all values of $\xi$. However, we do find that for $\xi >  \frac{m_h^2}{4m^2}$ and $ \lambda > g^2$, the real value of the pole mass is a real point inside the branch cut. This means that we cannot achieve the usual differentiation around this point. As a consequence, we cannot consistently construct the residue, so that we are unable to obtain a first-order spectral function. For the gauge field, we find the same problem when $\xi < \frac{1}{4}$.
	
	The foregoing mathematically correct observations clearly show that there is something physically wrong with using the elementary fields' spectral functions.  Luckily, all of these shortcomings are surpassed by using the gauge invariant composite operators.

	\subsection{Spectral properties for the composite fields \label{compp}}
\begin{figure}[H]
		\centering
		\includegraphics[width=18cm]{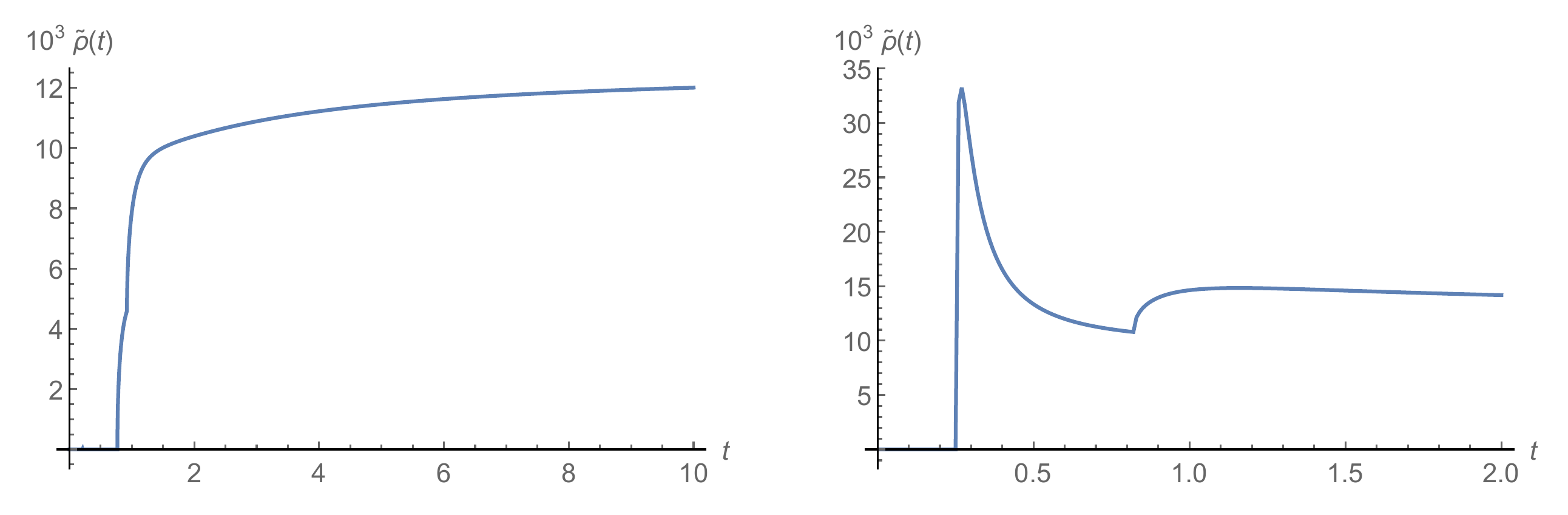}
		\caption{Spectral function for the two-point correlation function $\langle {O}(p){O}(-p) \rangle$, with $t$ given in units of $\mu^2$, for the Region I (left) and Region II (right) with parameter values given in Table \ref{tabel}. }
		\label{Y}
	\end{figure}

	For the scalar composite operator ${O}(x)$, whose two-point function is given by expression \eqref{WW}, we find the first-order pole mass for Region I
	\beq
	m_{OO, \rm pole}^2 &=& 0.207\, \mu^2, \label{OOa}
	\eeq
	and for Region II
	\beq
	m_{OO, \rm pole}^2 &=& 0.206 \, \mu^2, \label{OOb}
	\eeq
	which is equal to the pole mass of the elementary Higgs field in \eqref{higgsmass}, as we expect from eq.~\eqref{zlf}. Following the steps from Section~\ref{2a}, we find the first-order residue
	\beq
	Z= 1.11 \, v^2
	\eeq
	for Region I and
	\beq
	Z=1.01 \,v^2
	\eeq
	for Region II. The first order spectral function for $\braket{O(x)\,O(y)}$ is shown in  FIG.~\ref{Y}. Comparing this result with that of the spectral function of the Higgs field in  FIG.~\ref{Y3}, we see a two-particle state for the Higgs field at $t=(m_h+m_h)^2$, and a two-particle state for the gauge vector field, starting at $t=(m+m)^2$. The difference is that for the gauge invariant correlation function $\braket{O(x)\,O(y)}$ we no longer have the unphysical Goldstone two-particle state. Due to the absence of this negative contribution, the spectral function is positive throughout the spectrum. In fact, we see that for bigger values of $\xi$, we find that the spectral function of the elementary Higgs field resembles more and more the spectral function of the composite operator $O(x)$. This makes sense, since for $\xi \rightarrow \infty$, we are approaching the unitary gauge which has a more direct link with the physical spectrum of the elementary excitations. In Appendix~\ref{un}, one finds a detailed discussion about the unitary gauge limit $\xi \rightarrow \infty$ as well as the calculation of the spectral function. Of course, the unitary gauge is not well-suited for computations at the quantum level due to the non-renormalizability issue in general, and non-curable overlapping divergences in particular, but for gauge invariant quantities, it was already noted in \cite{irges2017renormalization} that (for the $\beta$-functions) the unitary gauge choice can still be useful, at least up to one-loop order. Therefore, in Appendix~\ref{un}, we have analyzed, at one-loop order where there are no overlapping divergences yet, the scalar and vector correlation functions directly in the unitary gauge and we do find that their \emph{finite} pieces are identical to those obtained in the $R_\xi$-gauge discussed so far. Given that several propagators are trivial in the unitary gauge, the computations in the latter gauge are evidently far more economical. A similar observation was already made in the $U(1)$ case as well, see \cite{Capri:2020ppe}.

 The asymptotic (constant) behaviour is directly related to the (classical) dimension of the used composite operator.
	
	\subsubsection{The vector composite operator $R^a_\mu(x)$}
\begin{figure}[H]
		\centering
		\includegraphics[width=18cm]{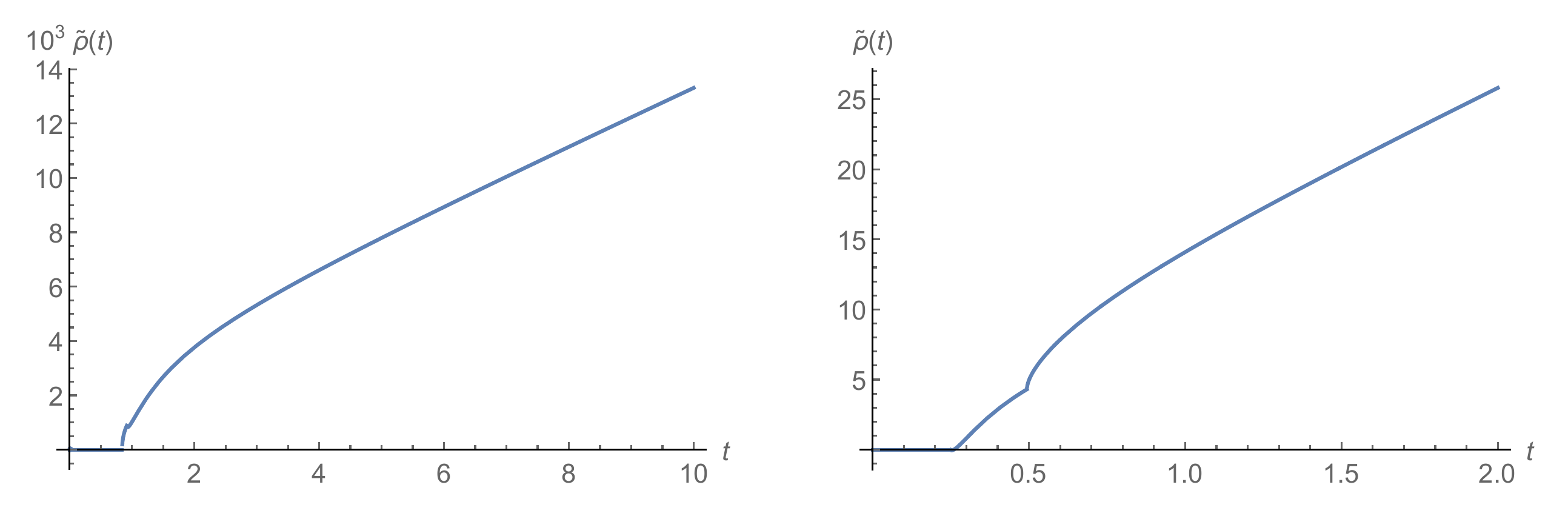}
		\caption{Spectral function for the transverse two-point correlation function  $ G_{R}(p^2)$, with $t$ given in units of $\mu^2$, for the Region I (left) and Region II (right) with parameter values given in Table \ref{tabel}. }
		\label{rtrt}
	\end{figure}

	For the transverse part of the two-point correlation function 	$ G_{R}(p^2)$, eq.~\eqref{dk3}, for our set of parameters we find the first-order pole mass: in Region I
	\beq
	m_{R, \rm pole}^2 &=& 0.274 \, \mu^2  \label{voa}
	\eeq
	and Region II
	\beq
	m_{R, \rm pole}^2 &=& 0.065 \, \mu^2  \label{vob}
	\eeq
	which is the same as the pole mass of the transverse gauge field, eq.~\eqref{polem}, in agreement with eq.~\eqref{zlf}. Following the steps from \ref{2a}, we find the first-order residue
	\beq
	Z=\frac{1}{16}g^2 v^4 \left(1.27\right)
	\eeq
	for Region I and
	\beq
	Z=\frac{1}{16}g^2 v^4 \left(1.05\right)
	\eeq
	for Region II. The first order spectral function for $ G_{R}(p^2)$  is shown in  FIG.~\ref{rtrt}. Comparing this to the spectral function of the gauge vector field in  FIG.~\ref{Y3}, we see a two-particle state at $t=(m_h+m)^2$, and a two-particle state for the gauge field, starting at $t=(m+m)^2$. Again, as in the case of the two-point correlation function of the scalar operator $O(x)$, the difference is that for this gauge invariant correlation function we no longer have the unphysical Goldstone/gauge boson two-particle state. Due to the absence of this negative contribution, the spectral function is positive throughout the spectrum. In fact, we see that for bigger values of $\xi$, we find that the spectral function of the elementary gauge field resembles more and more the spectral function of the composite operator $R^a_{\mu}(x)$. As already mentioned previously, this relies on the fact that in the limit $\xi \rightarrow \infty$ we are approaching  the unitary gauge, see  Appendix \ref{un}.  Also here, the linear increase at large $t$ follows from the operator dimension.
	
	\section{Conclusion and outlook \label{VI}}
	
	This  work is the natural extension of  previous studies \cite{Dudal:2019aew, Dudal:2019pyg,Capri:2020ppe}, where the Abelian $U(1)$ Higgs model has been scrutinized by employing two local composite BRST invariant operators \cite{tHooft:1980xss,Frohlich:1980gj, Frohlich:1981yi}, whose two-point correlation functions provide a fully gauge independent description of the elementary excitations of the model, namely the Higgs and the massive gauge boson.
	
	This formulation generalizes to the non-Abelian Higgs model as, for example, the $SU(2)$ Yang--Mills theory with a single Higgs in the fundamental representation \cite{tHooft:1980xss,Frohlich:1980gj, Frohlich:1981yi}. This is the model which has been considered in the present analysis. The local gauge and BRST invariant composite operators $(O(x), R^a_\mu(x))$ which generalize their $U(1)$ counterparts are given in eq.~\eqref{scopt} and in eqs.~\eqref{threeop},\eqref{threeop1}.
	
	The two-point correlation functions $\langle O(x) O(y) \rangle$ and
	$\langle R^a_\mu(x) R^b_\nu(y) \rangle^T$, where the superscript $T$ stands for the transverse component, have been evaluated at one-loop order in the $R_\xi$-gauge and compared with the corresponding correlation functions of the elementary fields $\langle h(x) h(y) \rangle $ and $\langle A^a_\mu(x) A^b_\nu(y) \rangle^T$. It turns out that both $\langle O(x) O(y) \rangle$ and $\langle h(x) h(y) \rangle $ share the same gauge independent pole mass, eqs.\eqref{higgsmass},\eqref{higgsmass2},\eqref{OOa},\eqref{OOb}, in agreement with both Nielsen identities \cite{Nielsen:1975fs,Piguet:1984js,gambino1999fermion,gambino2000nielsen,Grassi:2000dz,Aitchison:1983ns,Andreassen:2014eha} and the BRST invariant nature of $O(x)$. Nevertheless, unlike the residue and spectral function of the elementary correlator $\langle h(x) h(y) \rangle $, which exhibit a strong unphysical dependence from the gauge parameter $\xi$, FIG.~\ref{Y33}, the spectral density of $\langle O(x) O(y) \rangle$ turns out to be $\xi$-independent and positive over the whole $p^2$ axis, FIG.~\ref{Y}.  The same features hold for $\langle A^a_\mu(x) A^b_\nu(y) \rangle^T$ and $\langle R^a_\mu(x) R^b_\nu(y) \rangle^T$. Again, both correlation function share the same $\xi$-independent pole mass, eqs.~\eqref{polema},\eqref{polema},\eqref{voa},\eqref{vob}. Though, unlike the $\xi$-dependent spectral function associated to $\langle A^a_\mu(x) A^b_\nu(y) \rangle^T$, FIG.~\ref{Y3}, that corresponding to $\langle R^a_\mu(x) R^b_\nu(y) \rangle^T$, FIG.~\ref{rtrt},  turns out to be independent from the gauge parameter $\xi$ and positive. As such, the local composite operators $(O(x), R^a_\mu(x))$ provide a fully BRST consistent description of the observable scalar (Higgs) and vector boson particles.
	
	It is worth mentioning here that, besides the BRST invariance of the gauge fixed action, the model exhibits an additional global custodial symmetry, eqs.\eqref{custodial},\eqref{beta}, according to which all fields carrying the index $a=1,2,3$, {\it i.e.}~$(A^a_\mu, b^a, c^a, {\bar c}^a, \rho^a)$, undergo a global transformation in the adjoint representation of $SU(2)$. The same feature holds for the composite operators $(O(x), R^a_\mu(x))$ which transform exactly as $h$ and $A^a_\mu$. More precisely, the operator $O(x)$ is a singlet under the custodial symmetry, while the operators $R^a_\mu$ transform like a triplet, eq.~\eqref{tripr}, so that the correlation function $\langle R^a_\mu(p) R^b_\nu(-p) \rangle$ displays the same $SU(2)$ structure of the elementary two-point function  $\langle A^a_\mu(p) A_\nu(-p)\rangle$, eqs.\eqref{cc1}-\eqref{cc4}. Although not being the aim of the present analysis, we expect that the existence of a global custodial symmetry will imply far-reaching consequences for the renormalization properties of the composite operators $(O(x), R^a_\mu(x))$ encoded in the corresponding Ward identities,  allowing a generalization of the $U(1)$ renormalizability analysis of \cite{Capri:2020ppe}.
	
	The analysis of the spectral properties of the operators $(O(x), R^a_\mu(x))$ worked out here could pave the route for more ambitious projects which might lead to an interesting interplay with possible future investigation on the lattice of the correlators $\langle O(x) O(y) \rangle$ and $\langle R^a_\mu(x) R^b_\nu(y) \rangle^T$, as already mentioned in \cite{maas2015field, Maas:2017wzi}. The BRST  invariant nature of $(O(x), R^a_\mu(x))$ makes them natural candidates to attempting at facing the  challenge of investigating the infrared non-perturbative behaviour of the model, trying to make contact with the  analytical lattice predictions of Fradkin and Shenker \cite{Fradkin:1978dv}, see also \cite{Greensite:2011zz}
	for a general overview and \cite{Cherman:2020hbe} for a new take on these matters.. This study would enable us to shed some light on the issue of the positivity of the spectral densities in the non-perturbative region, a topic which is currently under intensive investigation in confining  Yang--Mills theories without the presence of Higgs fields , see  \cite{Oehme:1979ai,cornwall1982dynamical,zwanziger1989local,Alkofer:2000wg,Cucchieri:2004mf,bowman2007scaling,Cucchieri:2007md,Aguilar:2008xm,Dudal:2008sp,Maas:2008ri,Fischer:2008uz,Binosi:2009qm,Bogolubsky:2009dc,Bornyakov:2009ug,Cucchieri:2009zt,tissier2010infrared,tissier2011infrared,Boucaud:2011ug,Serreau:2012cg,Oliveira:2012eh,cornwall2013positivity,Aguilar:2014tka,Pelaez:2014mxa,Reinosa:2014zta,Siringo:2016jrc,Cyrol:2016tym,Duarte:2016iko,Reinosa:2016iml,Frasca:2015yva,Lowdon:2017gpp,Comitini:2017zfp,Huber:2018ned,hayashi2018complex,Binosi:2019ecz,Gracey:2019xom,Kondo:2019rpa,Dudal:2019gvn,Fischer:2020xnb}. Finally, it would be worth to investigate to which extent the BRST invariant correlation functions $\langle O(x) O(y) \rangle$, $\langle R^a_\mu(x) R^b_\nu(y) \rangle^T$ could be affected by the existence of the non-perturbative effect of the Gribov copies, by means of the recent BRST invariant formulation of the (Refined) Gribov-Zwanziger horizon function
	\cite{Capri:2015ixa,Capri:2018ijg}.
	
	Another most promising avenue to further explore is the $SU(2)\times U(1)$ setting of the electroweak theory, where the gauge invariant description of electric charge will necessitate the combination of the (local) gauge invariant composite operators set out here and (non-local) dressed gauge invariant operators, see \cite{Lavelle:1995ty}.

	\section*{Acknowledgements}
	The authors would like to thank the Brazilian agencies CNPq and FAPERJ for financial support. This study was financed in part by the Coordena{\c c}{\~a}o de Aperfei{\c c}oamento de Pessoal de N{\'\i}vel Superior - Brasil (CAPES) - Finance Code 001. This paper is also part of the project INCT-FNA Process No. 464898/2014-5.
	S.P. Sorella is a level $1$ CNPq researcher under the contract $301030/2019-7$.  M.S. Guimaraes is a level 2 CNPq researcher under the contract 307801/2017-9 and L.F. Palhares is a level 2 CNPq researcher under the contract 311751/2019-9.

	\begin{appendix}
		
		\section{Propagators and vertices \label{appA}}
		The tree level elementary propagators of the fields are easily computed, for example by coupling a source $J_\varphi$ to each field $\varphi$ and computing at leading order the functional derivative $\frac{\delta^2 Z^c}{\delta J_\varphi(-p) \delta J_\varphi(p)}$ with $Z^c$ the generating functional of connected two-point functions. This leads to
		\beq \label{props}
		\langle A^a_{\m}(p)A^b_{\n}(-p)\rangle &=& \frac{\delta^{ab}}{p^2+ m^2} \mathcal{P}_{\m\n}(p)+\delta^{ab}\frac{\xi}{p^2 + \xi m^2}\mathcal{L}_{\m\n}(p),\nonumber\\
		\langle \r^a(p)\r^b(-p)\rangle &=&\frac{\delta^{ab}}{p^2 +\xi m^2},\nonumber\\
		\langle h(p)h(-p)\rangle &=&\frac{1}{p^2 + m_h^2},\nonumber\\
		\langle A^a_{\m}(p)b^b(-p)\rangle &=&\delta^{ab} \frac{p_{\m}}{p^2+\xi m^2},\nonumber\\
		\langle b^a (p) \rho^b (-k) \rangle &=& \delta^{ab}\frac{im}{p^2+ \xi m^2}
		\eeq
		and
		\beq
		\langle \bar{c}^a(p)c^b(-p)\rangle &=&\frac{\delta^{ab}}{p^2 +\xi m^2}
		\eeq
		for the ghost propagator. All other propagators are zero. This includes the mixed propagator, $\braket{A_{\mu}^a(x)\rho^b(y)}=0$, a well-known feature of the $R_{\xi}$-gauge \cite[Chap.~21]{Peskin:1995ev}.
			
		For all vertices, adopting the convention that the momentum is flowing towards the vertex, we get
		\begin{itemize}
			\item The $AAh$-vertex:
			$\Gamma_{A_{\mu}^a A_{\nu}^b h}(-p_1,-p_2,-p_3)=-\frac{g^2 v}{2}\delta_{\mu\nu}\delta^{ab} \delta(p_1+p_2+p_3)$.
			\item The $\rho\rho A$-vertex: $\Gamma_{\rho^a \rho^b A_{\mu}^c} (-p_1,-p_2,-p_3)= \frac{g}{2}i \epsilon^{abc}(p_{\mu,1}-p_{\mu,2})\delta(p_1+p_2+p_3)$.
			\item The $A\rho h$-vertex: $\Gamma_{A_{\mu}^a \rho^b h}(-p_1,-p_2,-p_3)=i\frac{g}{2}\delta^{ab}(p_{\mu,3}-p_{\mu,2})\delta(p_1+p_2+p_3)$.
			\item The $hhh$ vertex: $\Gamma_{hhh}(-p_1,-p_2,-p_3)=- 3 \l v\,\,\delta(p_1+p_2+p_3)$.
			\item The $h\rho \rho $ vertex: $\Gamma_{h\rho^a \rho^b}(-p_1,-p_2,-p_3)=-\lambda v \delta^{ab}\,\,\delta(p_1+p_2+p_3)$.
			\item The $AAA$-vertex: $\Gamma_{A_{\m}^a A_{\n}^bA_{\s}^c}(-p_1,-p_2,-p_3)=-igf^{abc}\left[(p_1-p_3)_{\nu}\delta_{\s\m}+(p_3-p_2)_{\mu}\delta_{\n\s}+(p_2-p_1)_{\s}\delta_{\n\m}\right]\delta(p_1+p_2+p_3)$.
			\item The $\overline c A c$-vertex: $\Gamma_{\overline c^a A_{\m}^bc^c}(-p_1,-p_2,-p_3)=ig f^{abc}p_{1,\m}\delta(p_1+p_2+p_3)$.
			\item The $AAAA$-vertex: $\Gamma_{A_{\mu}^{a}A_{\nu}^{b}A_{\rho}^{c}A_{\sigma}^{d}}(-p_1,-p_2,-p_3,-p_4)=g^{2}[f^{eab}f^{ecd}(\delta_{\mu\sigma}\delta_{\nu\rho}-\delta_{\mu\rho}\delta_{\nu\sigma})+f^{eac}f^{ebd}(\delta_{\mu\sigma}\delta_{\nu\rho}-\delta_{\mu\nu}\delta_{\rho\sigma})+f^{ead}f^{ebc}(\delta_{\mu\rho}\delta_{\nu\sigma}-\delta_{\mu\nu}\delta_{\rho\sigma})]\delta(p_1+p_2+p_3+p_4)$.
			\item The $AAhh$-vertex: $\Gamma_{A_{\m}^aA_{\n}^b hh}(-p_1,-p_2,-p_3,-p_4)= -\frac{1}{2}g^2\delta^{ab}\delta_{\m\n}$.
			\item The $AA\r\r$-vertex: $\Gamma_{A_{\m}^aA_{\n}^b \rho^c \rho^d} (-p_1,-p_2,-p_3,-p_4)= - \frac{1}{2}g^2 \delta_{\m\n} \delta^{ab}\delta^{cd}$.
		\end{itemize}
		
		\section{Custodial symmetry \label{appcust}}
		The custodial symmetry given in Section \ref{cust} is a result of the fact that the unbroken action is invariant under an $SU(2)_{\rm gauge} \times SU(2)_{\rm global}$ symmetry. This can be seen by writing $\Phi$ in the form of a  bi-doublet \cite{Shifman:2012zz}
		\beq
		\Phi= \begin{pmatrix}
			\phi^{0*} && \phi^+ \\
			-\phi^{+*} && \phi^0
		\end{pmatrix}.
		\eeq
		Clearly, the action,
		\beq
		\mathcal{L}_{\Phi}&=&\frac{1}{2} \text{Tr}\left(\mathcal{D}_{\mu} \Phi\right)^{\dagger}\left(\mathcal{D}_{\mu} \Phi\right)-\frac{\lambda}{4} \left( \text{Tr} \,\Phi^{\dagger} \Phi - v^2 \right)^2,
		\label{LL}
		\eeq
		is invariant under the $SU(2)_{\rm gauge} \times SU(2)_{\rm global}$ transformation
		\beq
		\Phi &\rightarrow& U(x) \Phi M^{-1}, \nonumber \\
		A_{\mu}&\rightarrow& U (x) A_{\mu} U^{-1}(x)+ \frac{i}{g}U(x)\partial_{\mu} U^{-1}(x), \label{bba}
		\eeq
		with $M$ an arbitrary $x$-independent matrix from $SU(2)_{\rm global}$. In the bidoublet notation, the expansion in eq.~\eqref{expansion2} becomes
		\beq
		\Phi &=& (h+v) \mathbf{1}+i \rho^a \tau^a,
		\eeq
		so the vev $\braket{\Phi}$ is not invariant under the transformation \eqref{bba}. However, it is invariant under the global diagonal $SU(2) \times SU(2)$ subgroup corresponding to $U(x)= M$. This is the custodial symmetry, given in infinitesimal form in eq.~\eqref{custodial}.
		
		\section{Elementary propagators in the $R_{\xi}$-gauge\label{appel}}
		Here we will calculate the one-loop corrections to the Higgs and gauge field propagator. This requires the calculation \footnote{We have used from \cite{passarino1979one} the technique of modifying integrals into ``master integrals'' without numerators.} of the Feynman diagrams as shown in FIGS.~\ref{more} and \ref{les}. To shorten the intermediate expressions, we will use the following functions:
		\beq\label{eta}
		\eta(m_1,m_2)&\equiv& \frac{1}{(4\pi)^{d/2}}\Gamma(2-\frac{d}{2})\int_0^1 dx \left(p^2 x(1-x)+x m_1+(1-x) m_2\right)^{d/2-2},\nonumber\\
		\chi(m_1)&\equiv&  \frac{1}{(4\pi)^{d/2}} \Gamma(1-\frac{d}{2}) m_1^{d/2-1}.
		\eeq
		Notice that the last four diagrams for both particles are zero for $\langle h \rangle=0$. In fact, including these diagrams has the same effect as making a shift in the vev of the scalar field $\Phi$  to demand $\langle h \rangle=0$, see the Appendix of \cite{Dudal:2019aew} for the technical details.  In the context of the FMS operators, we found it more convenient to expand around the (classical) $v$ that is gauge invariant, and thus to include the tadpoles.  Expanding the FMS operator around the quantum corrected vev would lead to cancellations in that quantum vev coming from the propagator loop corrections to render it gauge invariant again, indeed the minimum of the quantum corrected effective Higgs potential is not gauge invariant itself.
		
		\subsubsection{Higgs propagator \label{hp}}
		\begin{figure}[H]
			\centering
			\includegraphics[width=\linewidth]{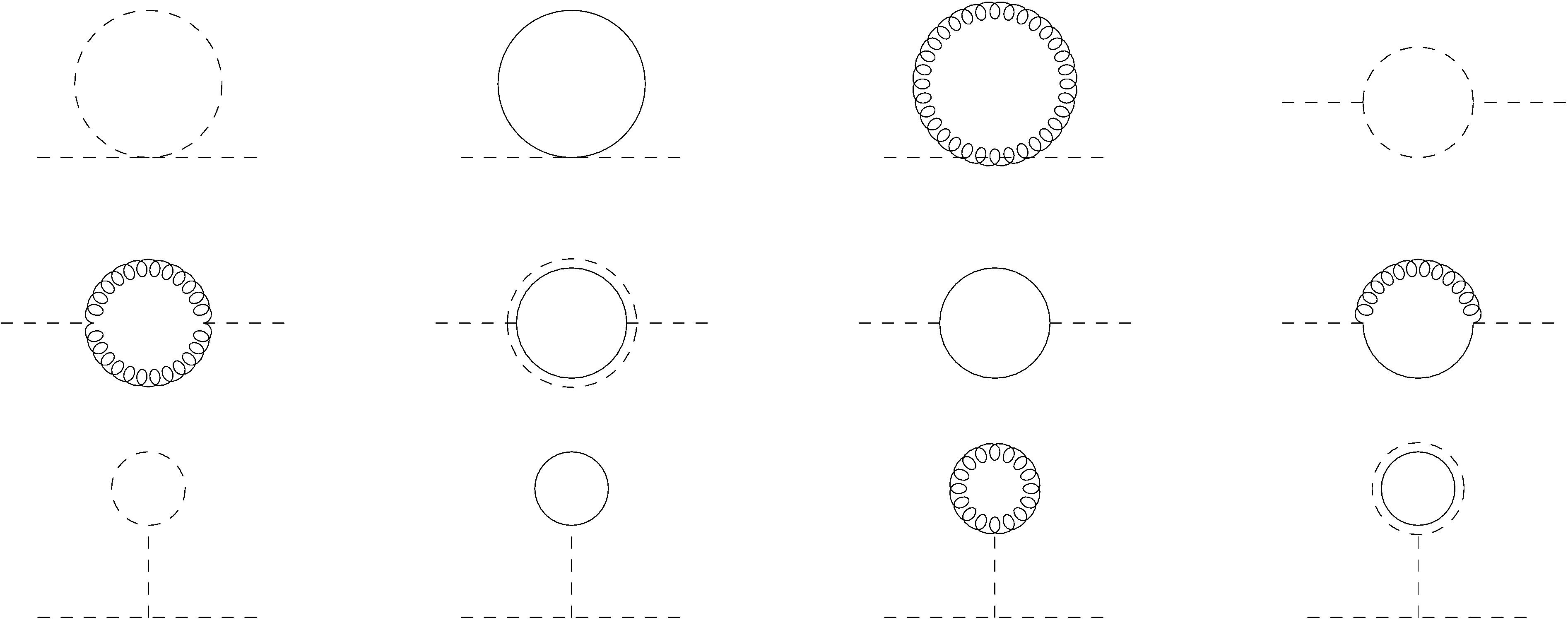}
			\caption{One-loop contributions to the propagator $\langle h(p)h(-p) \rangle$. Curly lines represent the gauge field, dashed lines the Higgs field, solid lines the Goldstone boson and double lines the ghost field.}
			\label{more}
		\end{figure}
		The first diagrams contributing to the Higgs self-energy are of the snail type, renormalizing the masses of the internal fields.
		The Higgs boson snail (first diagram in the first line of FIG.~\ref{more}):
		\beq
		\Gamma_{hh,1}(p^2)&=&-\frac{3 \lambda  \chi \left(m_h^2\right)}{2 \left(m_h^2+p^2\right)^2},
		\label{wat}
		\eeq
		the Goldstone boson snail  (second diagram in the first line of FIG.~\ref{more}):
		\beq
		\Gamma_{hh,2}(p^2) &=& -\frac{3 \lambda  \chi \left(m^2 \xi \right)}{2 \left(m_h^2+p^2\right)^2},
		\label{ii}
		\eeq
		and the gauge field  snail  (third diagram in the first line of FIG.~\ref{more}):
		\beq
		\Gamma_{hh,3}(p^2) &=& -\frac{3 (d-1) g^2 \chi \left(m^2\right)}{4 \left(m_h^2+p^2\right)^2}-\frac{3 g^2 \xi  \chi \left(m^2 \xi \right)}{4 \left(m_h^2+p^2\right)^2}.
		\eeq
		Next, we meet a couple of sunset diagrams. The Higgs boson sunset (fourth diagram in the first line of FIG.~\ref{more}):
		\beq
			\Gamma_{hh,4}(p^2)
		&=&\frac{9 \lambda ^2 v^2 \eta \left(m_h^2,m_h^2\right)}{2 \left(m_h^2+p^2\right)^2},
		\eeq
		the gauge field sunset (first diagram in the second line of FIG.~\ref{more}):
		\beq
		\Gamma_{hh,5}(p^2)	&=& \frac{3 g^2 \eta \left(m^2,m^2\right) \left(4 (d-1) m^4+4 m^2 p^2+p^4\right)}{8 m^2 \left(m_h^2+p^2\right)^2}+\frac{3 g^2 \left(2 m^2 \xi +p^2\right)^2 \eta \left(m^2 \xi ,m^2 \xi \right)}{8 m^2 \left(m_h^2+p^2\right)^2}\nonumber\\
		&-&\frac{3 g^2 \left(m^4 (\xi -1)^2+2 m^2 \xi  p^2+2 m^2 p^2+p^4\right) \eta \left(m^2,m^2 \xi \right)}{4 m^2 \left(m_h^2+p^2\right)^2}+\frac{3 g^2 (\xi -1) \chi \left(m^2\right)}{4 \left(m_h^2+p^2\right)^2}\nonumber\\
		&-&\frac{3 g^2 (\xi -1) \chi \left(m^2 \xi \right)}{4 \left(m_h^2+p^2\right)^2},
		\eeq
		the ghost sunset (second diagram in the second line of FIG.~\ref{more}):
		\beq
		\Gamma_{hh,6}(p^2)
		&=&-\frac{3 g^2 m^2 \xi ^2 \eta \left(m^2 \xi ,m^2 \xi \right)}{4 \left(m_h^2+p^2\right)^2},
		\eeq
		the Goldstone boson sunset (third diagram in the second line of FIG.~\ref{more}):
		\beq
		\Gamma_{hh,7}(p^2)&=&\frac{3 \lambda ^2 v^2 \eta \left(m^2 \xi ,m^2 \xi \right)}{2 \left(m_h^2+p^2\right)^2},
		\eeq
		and a mixed Goldstone-gauge sunset (fourth diagram in the second line of FIG.~\ref{more}):
		\beq
		\Gamma_{hh,8}(p^2)
		&=&\frac{3 g^2 \left(\left(m^2 (\xi -1)+p^2\right)^2+4 m^2 p^2\right) \eta \left(m^2,m^2 \xi \right)}{4 m^2 \left(m_h^2+p^2\right)^2}-\frac{3 g^2 \left(m^2 \xi +p^2\right)^2 \eta \left(m^2 \xi ,m^2 \xi \right)}{4 m^2 \left(m_h^2+p^2\right)^2}\nonumber\\
		&+&\frac{3 g^2 \chi \left(m^2 \xi \right) \left(m^2 (2 \xi -1)+p^2\right)}{4 m^2 \left(m_h^2+p^2\right)^2}-\frac{3 g^2 \chi \left(m^2\right) \left(m^2 (\xi -1)+p^2\right)}{4 m^2 \left(m_h^2+p^2\right)^2}.
		\eeq
		Finally, we have the tadpole diagrams. The Higgs tadpole (first diagram on the third line of  FIG.~\ref{more}):
		\beq
		\Gamma_{hh,9}(p^2) &=&\frac{9 \lambda ^2 v^2 \chi \left(m_h^2\right)}{2 m_h^2 \left(m_h^2+p^2\right)^2},
		\eeq
		the gauge tadpole  (second diagram on the third line of  FIG.~\ref{more}):
		\beq
		\Gamma_{hh,10}(p^2) &=& \frac{9 g \lambda  m v \left((d-1) \chi \left(m^2\right)+\xi  \chi \left(m^2 \xi \right)\right)}{2 m_h^2 \left(m_h^2+p^2\right)^2},
		\eeq
		the Goldstone boson tadpole  (third diagram on the third line of  FIG.~\ref{more}):
		\beq
		\Gamma_{hh,11}(p^2) &=& \frac{9 \lambda ^2 v^2 \chi \left(m^2 \xi \right)}{2 m_h^2 \left(m_h^2+p^2\right)^2},
		\label{uu}
		\eeq
		the ghost tadpole (fourth diagram on the third line of  FIG.~\ref{more}):	
		\beq
		\Gamma_{hh,12}(p^2)&=&-\frac{9 g \lambda  m \xi  v \chi \left(m^2 \xi \right)}{2 m_h^2}.
		\label{poi}
		\eeq
		Putting together eqs.~\eqref{wat} to \eqref{poi} we find the Higgs  up to first order in $\hbar$
		\beq
		\braket{h(x)\,h(y)}&=& \frac{1}{p^2+m_h^2}+g^2\Big\{ \frac{3  \left(4 (d-1) m^4+4 m^2 p^2+p^4\right)}{8 m^2}\eta \left(m^2,m^2\right)\nonumber\\
		&+&\frac{9 m_h^4}{8 m^2} \eta \left(m_h^2,m_h^2\right)+\frac{3 \left(m_h^4-p^4\right) }{8 m^2}\eta \left(m^2 \xi ,m^2 \xi \right)+\frac{ \left(6 (d-1) m^2-3 p^2\right)}{4 m^2}\chi \left(m^2\right)\nonumber\\
		&+&\frac{3 \left(m_h^2+p^2\right) }{4 m^2}\chi \left(m^2 \xi \right)+\frac{3 m_h^2}{4 m^2} \chi \left(m_h^2\right)\Big\} \frac{1}{(p^2+m_h^2)^2}.
		\label{hh1}
		\eeq

		\subsubsection{Gauge field propagator \label{Ap}}
		\begin{figure}[h]
			\centering
			\includegraphics[width=\textwidth]{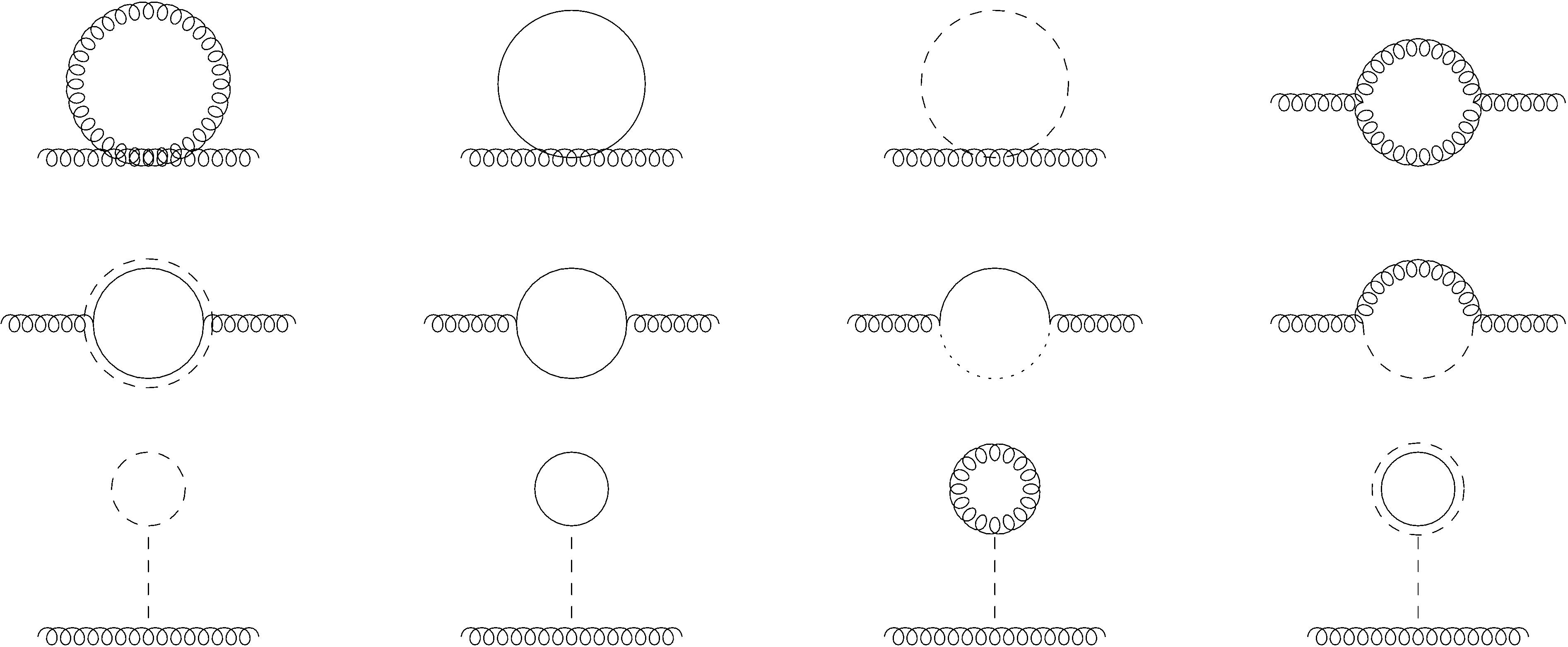}
			\caption{Contributions to the one-loop gauge field self-energy.}
			\label{les}
		\end{figure}

		The first diagram contributing to transverse part of the gauge field self-energy is the gauge field snail (first diagram in the first line of FIG.~\ref{les}) and gives a contribution:
		\beq
		\Pi_{AA^T, 1}(p^2)&=&\frac{2 g^2 \left(p^2-d \left(d^2-3 d+3\right) p^2\right) \chi \left(m^2\right)}{(d-1) d p^2 \left(m^2+p^2\right)^2}-\frac{2 g^2 \xi  \left((d-2) d p^2+p^2\right) \chi \left(m^2 \xi \right)}{(d-1) d p^2 \left(m^2+p^2\right)^2}.
		\label{besu}
		\eeq
		The second diagram is the Goldstone boson snail (second diagram in the first line of FIG.~\ref{les}):
		\beq
		\Pi_{AA^T, 2}(p^2)	&=&-\frac{3 g^2 \chi \left(m^2 \xi \right)}{4 \left(m^2+p^2\right)^2}.
		\eeq
		The third diagram is the Higgs boson snail (third diagram in the first line of FIG.~\ref{les}):
		\beq
		\Pi_{AA^T, 3}(p^2)	&=&-\frac{g^2 \chi \left(m_h^2\right)}{4 \left(m^2+p^2\right)^2}.
		\eeq
		The fourth diagram is the gauge field sunrise (fourth diagram in the first line of FIG.~\ref{les}):
		\beq
	\Pi_{AA^T, 4}(p^2) &=&g^2  \Big\{\frac{\eta \left(m^2,m^2 \xi \right) \left(2 m^2 p^2 (-2 d+\xi +3)+m^4 (\xi -1)^2+p^4\right)}{2 (d-1) m^4 p^2}\nonumber\\
		&-&\frac{\left(4 m^2+p^2\right) \eta \left(m^2,m^2\right) \left(4 (d-1) m^4+4 (3-2 d) m^2 p^2+p^4\right)}{4 (d-1) m^4 \left(m^2+p^2\right)^2}\nonumber\\
		&-&\frac{\left(4 m^2 \xi  p^4+p^6\right) \eta \left(m^2 \xi ,m^2 \xi \right)}{4 (d-1) m^4 \left(m^2+p^2\right)^2}\nonumber\\
		&+&\frac{\chi \left(m^2 \xi \right) \left(4 d^2 \left(m^2 (\xi +1) p^2+p^4\right)+d \left(m^4 (\xi -1)-m^2 (6 \xi +7) p^2+(\xi -7) p^4\right)+4 m^2 \xi  p^2\right)}{2 (d-1) d m^2 p^2 \left(m^2+p^2\right)^2}\nonumber\\
		&-&\frac{\chi \left(m^2\right) \left(4 d^2 p^4+d \left(m^4 (\xi -1)+m^2 (2 \xi -5) p^2+(\xi -7) p^4\right)+4 m^2 p^2\right)}{2 (d-1) d m^2 p^2 \left(m^2+p^2\right)^2}\Big\}.
		\eeq
		The fifth diagram is the ghost sunrise (first diagram in the second line of FIG.~\ref{les}):
		\beq
\Pi_{AA^T, 5}(p^2)
		&=&g^2 \left\{\eta \left(m^2 \xi ,m^2 \xi \right) \left(\frac{2  m^2 \xi }{(d-1) \left(m^2+p^2\right)^2}+\frac{ p^2}{2 (d-1) \left(m^2+p^2\right)^2}\right)-\frac{ \chi \left(m^2 \xi \right)}{(d-1) \left(m^2+p^2\right)^2}\right\}.
		\eeq
		The sixth diagram is the Goldstone sunrise (second diagram in the second line of FIG.~\ref{les}):
		\beq
	\Pi_{AA^T, 6}(p^2)
		&=&g^2 \left\{ \eta \left(m^2 \xi ,m^2 \xi \right) \left(-\frac{ m^2 \xi }{(d-1) \left(m^2+p^2\right)^2}-\frac{p^2}{4 (d-1) \left(m^2+p^2\right)^2}\right)+\frac{\chi \left(m^2 \xi \right)}{2 (d-1) \left(m^2+p^2\right)^2}\right\}.
		\eeq
		The seventh diagram is the mixed Goldstone-Higgs sunrise (third diagram in the third line of FIG.~\ref{les}):
		\beq
	\Pi_{AA^T, 7}(p^2)
		&=&g^2 \left\{ -\frac{\left(\left(-m_h^2+m^2 \xi +p^2\right)^2+4 m_h^2 p^2\right) \eta \left(m_h^2,m^2 \xi \right)}{4 (d-1) p^2 \left(m^2+p^2\right)^2}+\frac{\chi \left(m_h^2\right) \left(-m_h^2+m^2 \xi +p^2\right)}{4 (d-1) p^2 \left(m^2+p^2\right)^2}\right.\nonumber\\
		&+&\left.\frac{\chi \left(m^2 \xi \right) \left(m_h^2-m^2 \xi +p^2\right)}{4 (d-1) p^2 \left(m^2+p^2\right)^2}\right\}.
		\eeq
		The eighth diagram is the mixed Higgs-gauge field sunrise (fourth diagram in the second line of FIG.~\ref{les}):
		\beq
		\Pi_{AA^T, 8}(p^2)
		&=&g^2 \int_0^1 dx\frac{\left(\left(m_h^2-m^2 \xi +p^2\right)^2+4 m^2 \xi  p^2\right) \eta \left(m_h^2,m^2 \xi \right)}{4 (d-1) p^2 \left(m^2+p^2\right)^2}-\frac{\eta \left(m^2,m_h^2\right) \left(\left(m_h^2-m^2+p^2\right)^2-4 (d-2) m^2 p^2\right)}{4 (d-1) p^2 \left(m^2+p^2\right)^2}\nonumber\\
		&-&\frac{m^2 (\xi -1) \chi \left(m_h^2\right)}{4 (d-1) p^2 \left(m^2+p^2\right)^2}-\frac{\chi \left(m^2 \xi \right) \left(m_h^2-m^2 \xi +p^2\right)}{4 (d-1) p^2 \left(m^2+p^2\right)^2}+\frac{\chi \left(m^2\right) \left(m_h^2-m^2+p^2\right)}{4 (d-1) p^2 \left(m^2+p^2\right)^2}.
		\eeq
		Finally, we have four tadpole diagrams. The Higgs boson tadpole (first diagram of the third line in  FIG.~\ref{les}):
		\beq
		\Pi_{AA^T, 5}(p^2)
		&=& \frac{3 g m \chi \left(m_h^2\right)}{2 v \left(m^2+p^2\right)^2},
		\eeq
		the Goldstone boson tadpole (second diagram of the last line in  FIG.~\ref{les}):
		\beq
		\Pi_{AA^T, 6}(p^2)
		&=&\frac{3 g \lambda  m v \chi \left(m^2 \xi \right)}{2 m_h^2 \left(m^2+p^2\right)^2},
		\eeq
		The gauge field tadpole (third diagram of the last line in  FIG.~\ref{les}):
		\beq
		\Pi_{AA^T,7}(p^2)&=&\frac{3 (d-1) g^2 m^2 \chi \left(m^2\right)}{2 m_h^2 \left(m^2+p^2\right)^2}+\frac{3 g^2 m^2 \xi  \chi \left(m^2 \xi \right)}{2 m_h^2 \left(m^2+p^2\right)^2}
		\eeq
		and finally, the ghost tadpole (fourth diagram of the last line in  FIG.~\ref{les}):
		\beq
		\Gamma_{AA^T,8}(p^2)&=&-\frac{3 g^2 m^2 \xi  \chi \left(m^2 \xi \right)}{2 m_h^2 \left(m^2+p^2\right)^2}.
		\label{absu}
		\eeq
		Combining all these contributions \eqref{besu}-\eqref{absu}, we find the total one-loop correction to the gauge field self-energy
		\beq
		\braket{A^a_{\mu}(p) A^b_{\nu}(p)}^T&=&\frac{\delta^{ab}}{p^2+m^2}+\delta^{ab}g^2  \Big\{-\frac{ \left(2 (3-2 d) m^2 p^2+m_h^4-2 m_h^2 \left(m^2-p^2\right)+m^4+p^4\right)}{4 (d-1) p^2}\eta \left(m^2,m_h^2\right)\nonumber\\
		&+&\frac{\left(m^2+p^2\right)^2 \left(2 m^2 p^2 (-2 d+\xi +3)+m^4 (\xi -1)^2+p^4\right)}{2 (d-1) m^4 p^2}\eta \left(m^2,m^2 \xi \right) \nonumber\\
		&+&\frac{\left(m^4-p^4\right) \left(4 m^2 \xi +p^2\right) }{4 (d-1) m^4}\eta \left(m^2 \xi ,m^2 \xi \right)-\frac{\left(4 m^2+p^2\right)  \left(4 (d-1) m^4+4 (3-2 d) m^2 p^2+p^4\right)}{4 (d-1) m^4}\eta \left(m^2,m^2\right)\nonumber\\
		&+&\frac{\left(m_h^2 \left(-m^2 p^2 \left(8 d^2-24 d+4 \xi +13\right)-2 p^4 (4 d+\xi -7)+m^4 (1-2 \xi )\right)+6 (d-1)^2 m^4 p^2+m_h^4 m^2\right)}{4 (d-1) m_h^2 m^2 p^2}\chi \left(m^2\right) \nonumber\\
		&-&\frac{ \left((d-2) p^2+m_h^2-m^2\right)}{4 (d-1) p^2}\chi \left(m_h^2\right)+\frac{ \left(m^2 p^2 (5 d+4 \xi -13)+2 p^4 (4 d+\xi -7)+2 m^4 (\xi -1)\right)}{4 (d-1) m^2 p^2}\chi \left(m^2 \xi \right)\nonumber\\
		&+&\frac{3 }{4}\chi \left(m_h^2\right)+\frac{3}{4} \chi \left(m^2 \xi \right)\Big\}\frac{1}{(p^2+m^2)^2}
		\label{AA+}.
		\eeq

		\section{Contributions to $\langle O(p) O(-p) \rangle$ in the $R_{\xi}$-gauge \label{OO}}
		\begin{figure}[t]
			\centering
			\includegraphics[width=\linewidth]{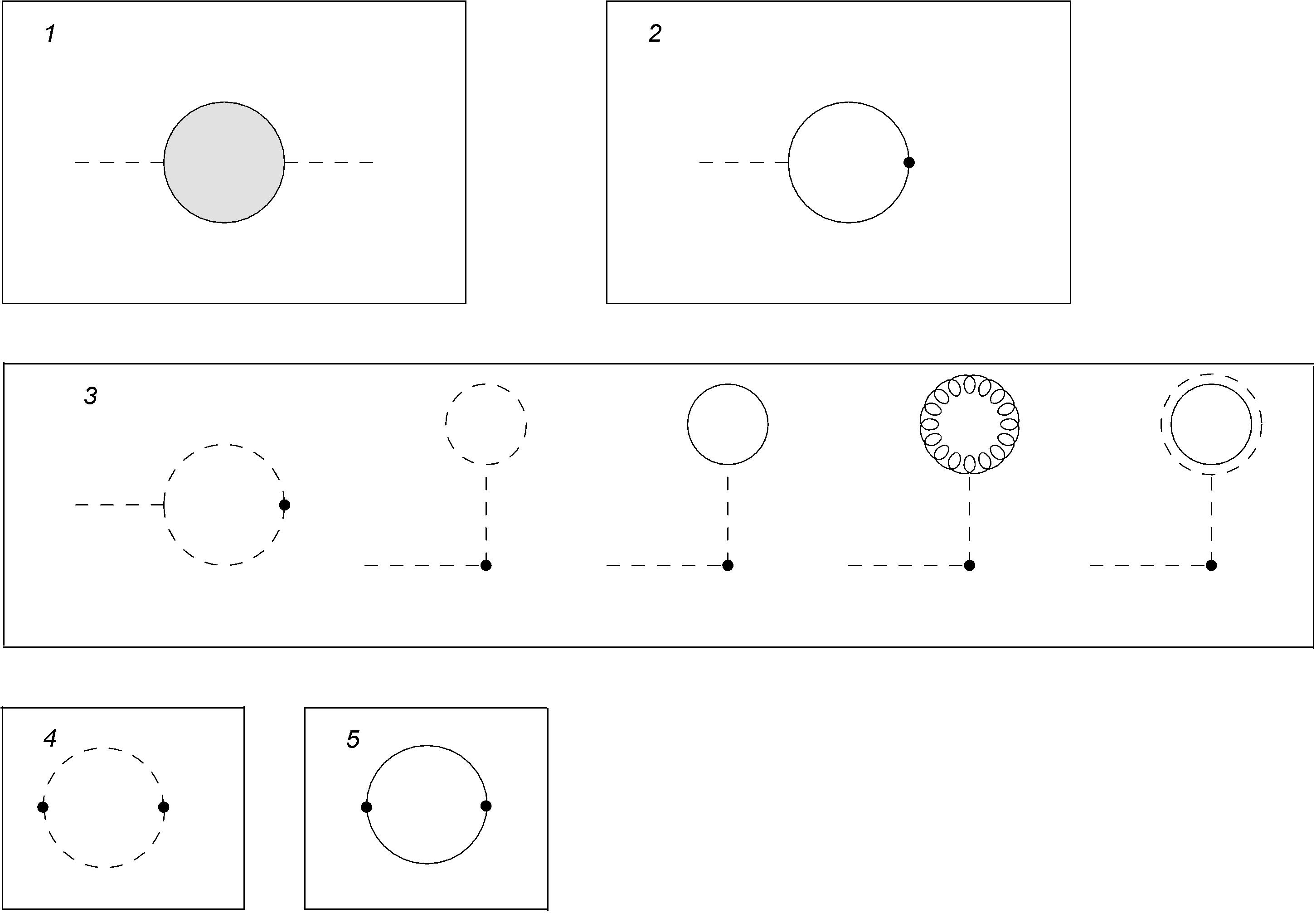}
			\caption{One-loop contributions for the propagator $\langle O(x)O(y) \rangle$. Curly lines represent the gauge field, dashed lines the Higgs field, solid lines the Goldstone boson and double lines the ghost field. The $\bullet$  indicates the insertion of a composite operator. }.  \label{hhhh}
		\end{figure}

		The diagrams which contribute to the correlation function $\langle O(p) O(-p) \rangle$ are depicted in FIG.~\ref{hhhh}. The first term (first box in FIG.~\ref{hhhh}) is $v^2$ times the one-loop correction to the Higgs propagator, given in  eq.~\eqref{hh1}. The second term (second box in FIG.~\ref{hhhh}) is
		\beq
		v \langle h(p) (\rho^a\rho^a)(-p) \rangle &=& -3\frac{m_h^2 \eta \left(m^2 \xi ,m^2 \xi \right)}{m_h^2+p^2}.
		\eeq
		The third term (third box in FIG.~\ref{hhhh}) is
		\beq
		v \langle h(p) h^2(-p) \rangle &=&-\frac{3 m_h^2 \eta \left(m_h^2,m_h^2\right)}{m_h^2+p^2}-\frac{3 \chi \left(m_h^2\right)}{m_h^2+p^2}-\frac{\chi \left(m^2 \xi \right)}{m_h^2+p^2}-\frac{2 (d-1) m^2 \chi \left(m^2\right)}{m_h^2 \left(m_h^2+p^2\right)}-\frac{2 m^2 \xi  \chi \left(m^2 \xi \right)}{m_h^2 \left(m_h^2+p^2\right)}\nonumber\\
		&+& \frac{2 m^2 \xi  \chi \left(m^2 \xi \right)}{m_h^2 \left(m_h^2+p^2\right)}.
		\eeq
		The fourth term (fourth box in FIG.~\ref{hhhh}) is
		\beq
		\langle m_h^2(p)m_h^2(-p) \rangle &=& \frac{1}{2}\eta \left(m_h^2,m_h^2\right).
		\eeq
		The fifth term (fifth box in FIG.~\ref{hhhh}) is
		\beq
		\langle (\rho^a\rho^a)(p) (\rho^b\rho^b)(-p) \rangle &=& \frac{3}{2} \eta \left(\xi m^2,\xi m^2\right)
		\eeq
		and together these terms give the correlation function of the scalar composite operator $O(p)$ up to first order in $\hbar$
		\beq
		\langle O(p) O(-p) \rangle &=& \frac{v^2}{p^2+m_h^2}+ \Big\{\frac{3}{2}\eta \left(m^2,m^2\right)(4 (d-1) m^4+4 m^2 p^2+p^4)+\frac{1}{2}(p^2-2 m_h^2)^2\eta \left(m_h^2,m_h^2\right)\nonumber\\
		&-&\frac{3 p^2 \chi (m^2) (2 (d-1) m^2+m_h^2)}{m_h^2}-3 p^2 \chi  (m_h^2)\Big\}\frac{1}{(p^2+m_h^2)^2}. \label{ch}
		\eeq

\section{Contributions to $\langle R_\mu^a(p) R_\nu^a(-p) \rangle$ in the $R_{\xi}$-gauge \label{RRR}}

 	\begin{figure}[t]
 	\centering
 	\includegraphics[width=\linewidth]{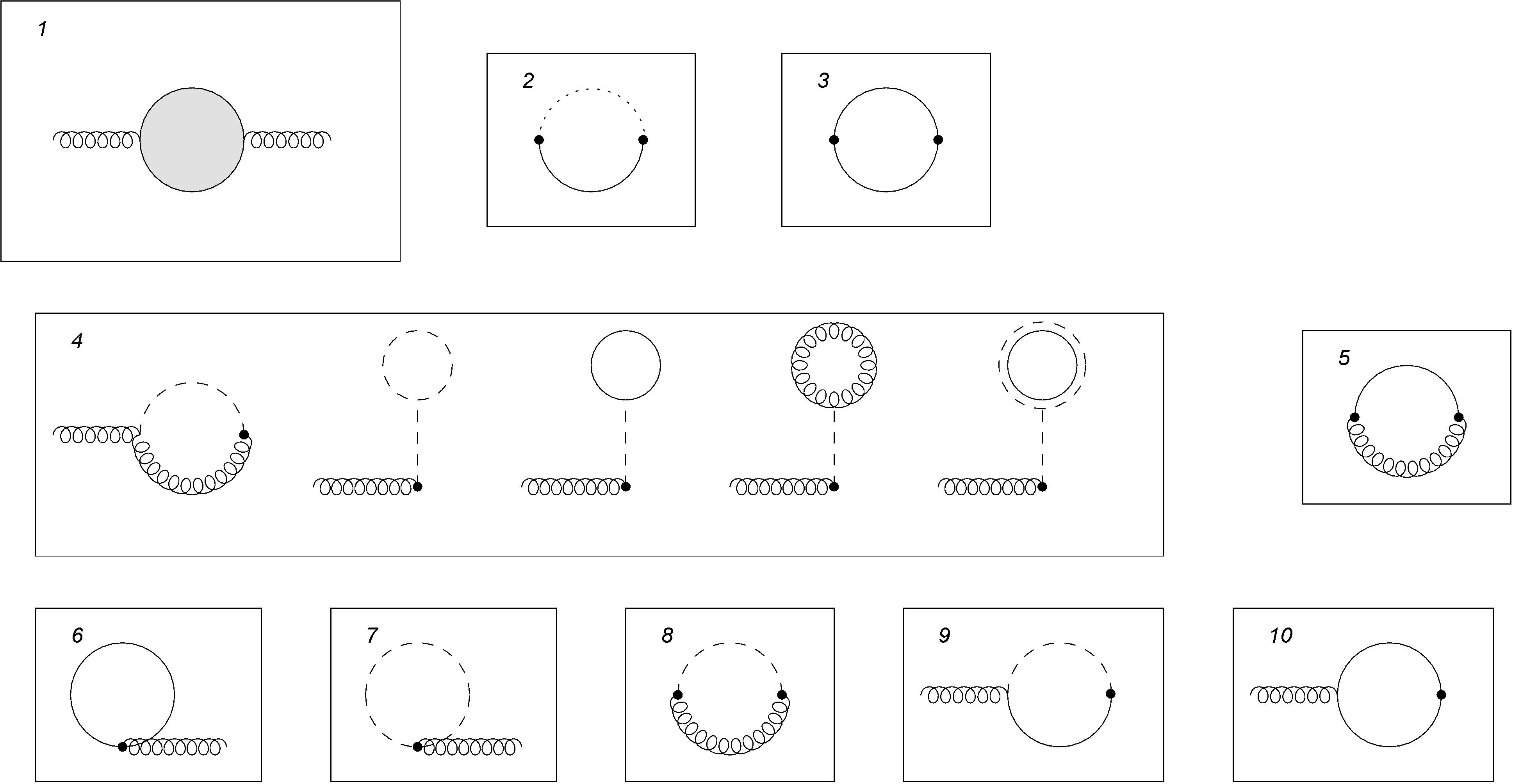}
 	\caption{One-loop contributions for the propagator $\langle R^a_{\mu}(x)R^a_{\nu}(y) \rangle$ in the unitary gauge. Curly lines represent the gauge field, dashed lines the Higgs field, solid lines the Goldstone boson and double lines the ghost field. The $\bullet$  indicates the insertion of a composite operator. }.  \label{AAAA}
 \end{figure}

The diagrams which contribute to the correlation function $\langle R_{\mu}^a(p) R_{\nu}^a(-p) \rangle$ are depicted in FIG.~\ref{AAAA}. The first term (first box in FIG.~\ref{AAAA}) is $\frac{1}{16}g^2 v^4$ times the one-loop corrected function, given by eq. \eqref{AA+}. The second term (second box in FIG.~\ref{AAAA}) is
\beq
	-\left\langle\rho^{a}(p) \partial_{\mu} h(p), \partial_{\nu} \rho^{a}(-p) h(-p)\right\rangle &=&\frac{3}{4(d-1)p^2} \Bigg\{\chi\left(m^{2} \xi\right)\left(m_h^{2}+ p^{2}-m^{2} \xi\right)\nonumber -\left(4 m_h^2p^2+\left(-m_h^{2}+m^{2} \xi+p^{2}\right)^{2}\right) \eta\left(m_h^{2}, m^{2} \xi\right)\\
	&& \,
-\chi\left(m_h^{2}\right)\left(m_h^{2}-m^{2} \xi-p^{2}\right)\Bigg\} \mathcal{P}_{\mu \nu}(p).
\eeq
The third term (third box in FIG.~\ref{AAAA}) is
\beq
-\frac{1}{2}\left\langle\rho^{a}(x) \partial_{\mu} \rho^{b}(x), \partial_{\nu} \rho^{a}(y) \rho^{b}(y)\right\rangle=\left\{\frac{3 \chi\left(m^{2} \xi\right)}{2(d-1)}-\frac{3\left(4 m^{2} \xi+p^{2}\right) \eta\left(m^{2} \xi, m^{2} \xi\right)}{4(d-1)}\right\} \mathcal{P}_{\mu \nu}(p).
\eeq 	
The fourth term (fourth box in FIG.~\ref{AAAA})is
\beq
	\frac{1}{4} g^{2} v^{3}\left\langle A_{\mu}^{a}(x), A_{\nu}^{a}(y) h(y)\right\rangle &=&\left\{-\frac{3 m^{2}\left(\left(-m_h^{2}+m^{2} \xi+p^{2}\right)^{2}+4 m_h^{2} p^{2}\right) \eta\left(m_h^{2}, m^{2} \xi\right)}{2(d-1) p^{2}\left(m^{2}+p^{2}\right)}\right.\nonumber\\
	&& +\frac{3 m^{2} \eta\left(m^{2}, m_h^{2}\right)\left(4 p^{2}\left(m_h^{2}-(d-1) m^{2}\right)+\left(-m_h^{2}+m^{2}+p^{2}\right)^{2}\right)}{2(d-1) p^{2}\left(m^{2}+p^{2}\right)}\nonumber \\
	&& +\frac{3 m^{2} \chi\left(m^{2} \xi\right)\left(\frac{m_h^{2}-m^{2} \xi+p^{2}}{(d-1) p^{2}}-3\right)}{2\left(m^{2}+p^{2}\right)}+\frac{3 m^{4} \chi\left(m_h^{2}\right)\left(\frac{\xi-1}{(d-1) p^{2}}-\frac{3}{ m^{2}}\right)}{2\left(m^{2}+p^{2}\right)} \nonumber\\
	&& \left.-\frac{3 m^{2} \chi\left(m^{2}\right)\left(6(d-1)^{2} m^{2} p^{2}+m_h^{4}+m_h^{2}\left(p^{2}-m^{2}\right)\right)}{2(d-1) m_h^{2} p^{2}\left(m^{2}+p^{2}\right)}\right\} \mathcal{P}_{\mu \nu}(p).
\eeq
The fifth term (fifth box in FIG.~\ref{AAAA}) is
\beq
\frac{1}{6} g^{2} v^{2}\left\langle\rho^{a}(x) A_{\mu}^{b}(x), \rho^{a}(y) A_{\nu}^{b}(y)\right\rangle&=&\Bigg\{
-\frac{3  \left(\left(m^2 (\xi -1)+p^2\right)^2-4 (d-2) m^2 p^2\right)\eta \left(m^2,m^2 \xi \right)}{2 (d-1) p^2}\nonumber\\
&+&\frac{3 \left(4 m^2 \xi +p^2\right) \eta \left(m^2 \xi ,m^2 \xi \right)}{2 (d-1)}+\frac{3 \chi \left(m^2\right) \left(m^2 (\xi -1)+p^2\right)}{2 (d-1) p^2}\nonumber\\
&-&\frac{3 \chi \left(m^2 \xi \right) \left(m^2 (\xi -1)+p^2\right)}{2 (d-1) p^2}
\Bigg\} \mathcal{P}_{\mu \nu}(p)
\eeq
The sixth term is
\beq
-\frac{1}{24} g^{2} v^{2}\left\langle\rho^{a}(x) \rho^{a}(x) A_{\mu}^{b}(x), A^{b}(y)\right\rangle=-\frac{3 g^{2} v^{2} \chi\left(\xi m^{2}\right)}{8\left(m^{2}+p^{2}\right)} \mathcal{P}_{\mu \nu}(p).
\eeq
The seventh term (seventh box in FIG.~\ref{AAAA}) is
\beq
\frac{1}{8} g^{2} v^{2}\left\langle h(x) h(x) A_{\mu}^{a}(x), A_{\nu}^{a}(y)\right\rangle=\frac{3 g^{2} v^{2} \chi\left(m_h^{2}\right)}{8\left(m^{2}+p^{2}\right)} \mathcal{P}_{\mu \nu}(p).
\eeq
The eighth term (eighth box in FIG.~\ref{AAAA}) is
\beq
\frac{1}{4} g^{2} v^{2}\left\langle h(x) A_{\mu}^{a}(x), h(y) A_{\nu}^{a}(y)\right\rangle &=&3 m^{2} \left\{\frac{\left(\left(-m_h^{2}+m^{2} \xi+p^{2}\right)^{2}+4 m_h^{2} p^{2}\right) \eta\left(m_h^{2}, m^{2} \xi\right)}{4(d-1) m^{2} p^{2}}\right.\nonumber\\
&&-\frac{\eta\left(m^{2}, m_h^{2}\right)\left(4 p^{2}\left(m_h^{2}-(d-1) m^{2}\right)+\left(-m_h^{2}+m^{2}+p^{2}\right)^{2}\right)}{4(d-1) m^{2} p^{2}} \nonumber\\
&&+\frac{\chi\left(m^{2} \xi\right)\left(-m_h^{2}+m^{2} \xi-p^{2}\right)}{4(d-1) m^{2} p^{2}}+\frac{\chi\left(m^{2}\right)\left(m_h^{2}-m^{2}+p^{2}\right)}{4(d-1) m^{2} p^{2}} \nonumber\\
&&\left.-\frac{(\xi-1) \chi\left(m_h^{2}\right)}{4(d-1) p^{2}}\right\} \mathcal{P}_{\mu \nu}(p).
\eeq
The ninth term (ninth box in FIG.~\ref{AAAA}) is
\beq
\frac{1}{2} g v^{2}\left\langle\partial_{\mu} h(x) \rho^{a}(x), A_{\nu}^{a}(y)\right\rangle &=&-\frac{3}{2} g^2 v^{2} \left\{-\frac{\left(\left(-m_h^{2}+m^{2} \xi+p^{2}\right)^{2}+4 m_h^{2} p^{2}\right) \eta\left(m_h^{2}, m^{2} \xi\right)}{4(d-1) p^{2}\left(m^{2}+p^{2}\right)}\right.\nonumber\\
&&\left.+\frac{\chi\left(m_h^{2}\right)\left(-m_h^{2}+m^{2} \xi+p^{2}\right)}{4(d-1) p^{2}\left(m^{2}+p^{2}\right)}+\frac{ \chi\left(m^{2} \xi\right)\left(m_h^{2}-m^{2} \xi+p^{2}\right)}{4(d-1) p^{2}\left(m^{2}+p^{2}\right)}\right\} \mathcal{P}_{\mu \nu}(p).
\eeq
The tenth term (tenth box in FIG.~\ref{AAAA}) is
\beq
 -\frac{1}{4} g v^{2}\varepsilon^{abc}\left\langle A^{a}(x), \rho^{b} \partial_{\mu} \rho^{c}\right\rangle=-\frac{3}{4}g^2  v^{2} \left\{\frac{\left(4 m^{2} \xi p^{2}+p^{4}\right) \eta\left(m^{2} \xi, m^{2} \xi\right)}{2(d-1) p^{2}\left(m^{2}+p^{2}\right)}-\frac{ \chi\left(m^{2} \xi\right)}{(d-1)\left(m^{2}+p^{2}\right)}\right\}\mathcal{P}_{\mu \nu}(p)
\eeq
and together these terms give the transverse part of the correlation function of the scalar composite operator $R^a_{\mu}$ up to first order in $\hbar$
\beq
	\left\langle R_{\mu}^{a}(x), R_{\nu}^{a}(y)\right\rangle^{T} &=&\frac{3}{16} g^{2} v^{4} \frac{1}{p^{2}+m^{2}}-3 \left\{-\frac{p^{2}\left(2(3-2 d) m^{2} p^{2}-2 m_{h}^{2}\left(m^{2}-p^{2}\right)+m_{h}^{4}+m^{4}+p^{4}\right)}{4(d-1)} \eta\left(m^{2}, m_h^{2}\right)\right.\nonumber\\
	&&-\frac{\left(4 m^{2}+p^{2}\right)\left(4(d-1) m^{4}+4(3-2 d) m^{2} p^{2}+p^{4}\right)}{4(d-1)} \eta\left(m^{2}, m^{2}\right) \nonumber\\
	&&+\frac{\chi\left(m^{2}\right)\left(m_{h}^{2}\left(-8\left(d^{2}-3 d+2\right) m^{4}+(15-8 d) m^{2} p^{2}+3 p^{4}\right)-6(d-1)^{2} m^{4}\left(m^{2}+2 p^{2}\right)+p^{2} m_{h}^{4}\right)}{4(d-1) m_{h}^{2}} \nonumber \\
	&&\left.+\frac{\chi\left(m_{h}^{2}\right)\left(-2(d-1) m^{4}+(5-4 d) m^{2} p^{2}-p^{2} m_{h}^{2}+p^{4}\right)}{4(d-1)}\right\} \frac{1}{\left(m^{2}+p^{2}\right)^{2}}.
\eeq

		\section{ Fundamental Feynman integral \label{appfeyn}}
		\beq
		\int_0^1 dx \ln \frac{p^2 x(1-x)+x m_1^2+(1-x)m_2^2}{\mu^2}=-2&+&\frac{1}{2 p^2}\Bigg\{m_1^2 \ln(\frac{m_2^2}{m_1^2})+m_2^2 \ln(\frac{m_1^2}{m_2^2})+p^2 \ln(\frac{m_1^2 m_2^2}{\mu^4})\nonumber\\
		&-&2 \sqrt{-m_1^4+2 m_1^2 m_2^2-2 m_1^2 p^2-m_2^4-2 m_2^2 p^2-p^4}\nonumber\\
		&\times&\tan^{-1}\Big[\frac{-m_1^2+m_2^2-p^2}{\sqrt{-m_1^4+2 m_1^2 (m_2^2-p^2)-(m_2^2+p^2)^2}}\Big]\nonumber\\
		&+&2 \sqrt{-m_1^4+2 m_1^2 m_2^2-2 m_1^2 p^2-m_2^4-2 m_2^2 p^2-p^4}\nonumber\\
		&\times&\tan^{-1}\Big[\frac{-m_1^2+m_2^2+p^2}{\sqrt{-m_1^4+2 m_1^2 (m_2^2-p^2)-(m_2^2+p^2)^2}}\Big]\Bigg\}.\label{bigI}
		\eeq

		\section{A digression on the unitary gauge \label{un}}
It is well-known that in the unitary gauge the unphysical fields, like the Goldstone and ghost fields, decouple, a feature which allows for a more direct link with the spectrum of the elementary excitations of the model.  However, this gauge is known to be  non-renormalizable.  In fact, working directly with the elementary tree level propagators taken a priori in the unitary limit, {\it i.e.}~$\xi \rightarrow \infty$, and following the steps of dimensional regularization, we find that the divergent part of the inverse Higgs propagator  reads
		\beq
		G^{-1}_{hh, \rm div}(p^2)&=&\frac{3 g^2 \left(m_h^4+6 m^2 p^2+p^4\right)}{64 \pi ^2 m^2 \epsilon }.   \label{divu}
		\eeq
We used here that the surviving tree level propagators in the unitary gauge are
	\beq
	\langle A^a_{\m}(p)A^b_{\n}(-p)\rangle &=& \frac{\delta^{ab}}{p^2+ m^2} \mathcal{P}_{\m\n}(p)+\delta^{ab}\frac{1}{m^2}\mathcal{L}_{\m\n}(p)=  \frac{\delta^{ab}}{p^2+ m^2}\Big(\delta_{\mu \nu}+ \frac{p_{\m}p_{\n}}{m^2}\Big),\nonumber\\
	\langle h(p)h(-p)\rangle &=&\frac{1}{p^2 + m_h^2}\,
	\label{uuu}
	\eeq
	with all other propagators, {\it i.e.}~the Goldstone and Faddeev-Popov ghost propagators, vanishing. In expression \eqref{divu} we clearly see  the presence of the term  $\sim \frac{p^4}{\epsilon m^2}$, signalling the aforementioned issue of the non-renormalizability.  In essence, this can be traced back to the tree level presence of $\frac{p_{\m}p_{\n}}{m^2}$, a non power-counting controllable term.

\begin{figure}[H]
			\centering
			\includegraphics[width=18cm]{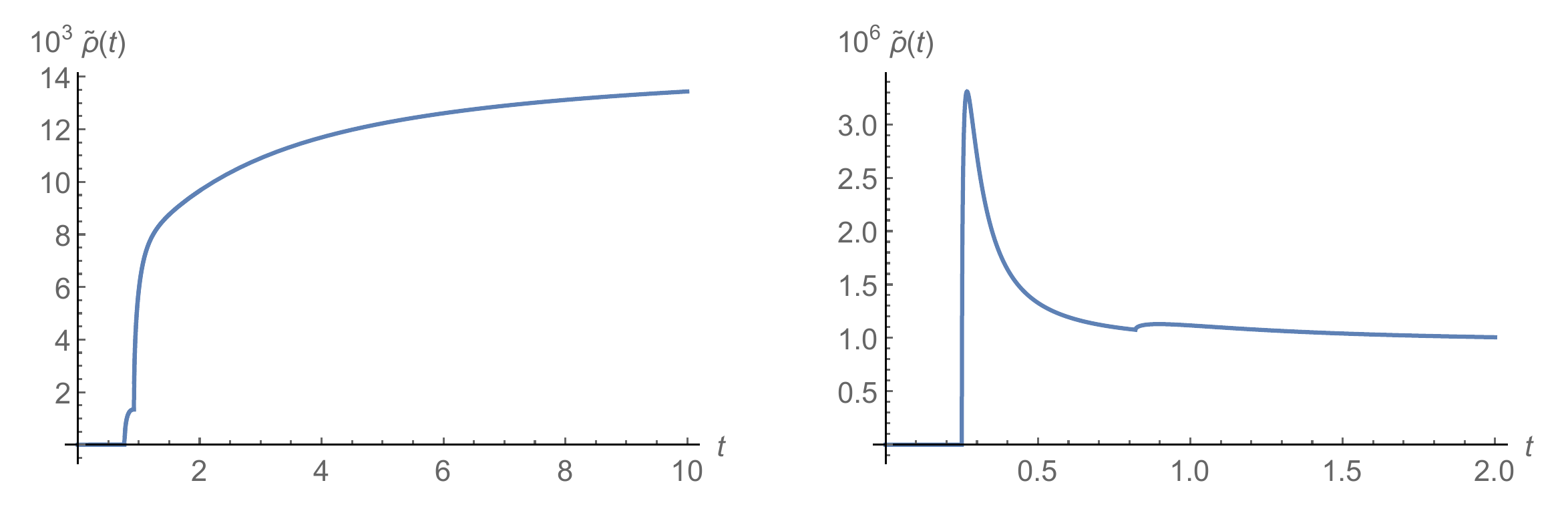}
			\caption{Spectral function for the propagator $\langle h(p) h(-p) \rangle$ in the unitary gauge, with $t$ given in unity of $\mu^2$, for the Region I (left) and Region II (right), with parameter values given in Table \ref{tabel}.}
			\label{Yhh}
		\end{figure}

\begin{figure}[H]
			\centering
			\includegraphics[width=18cm]{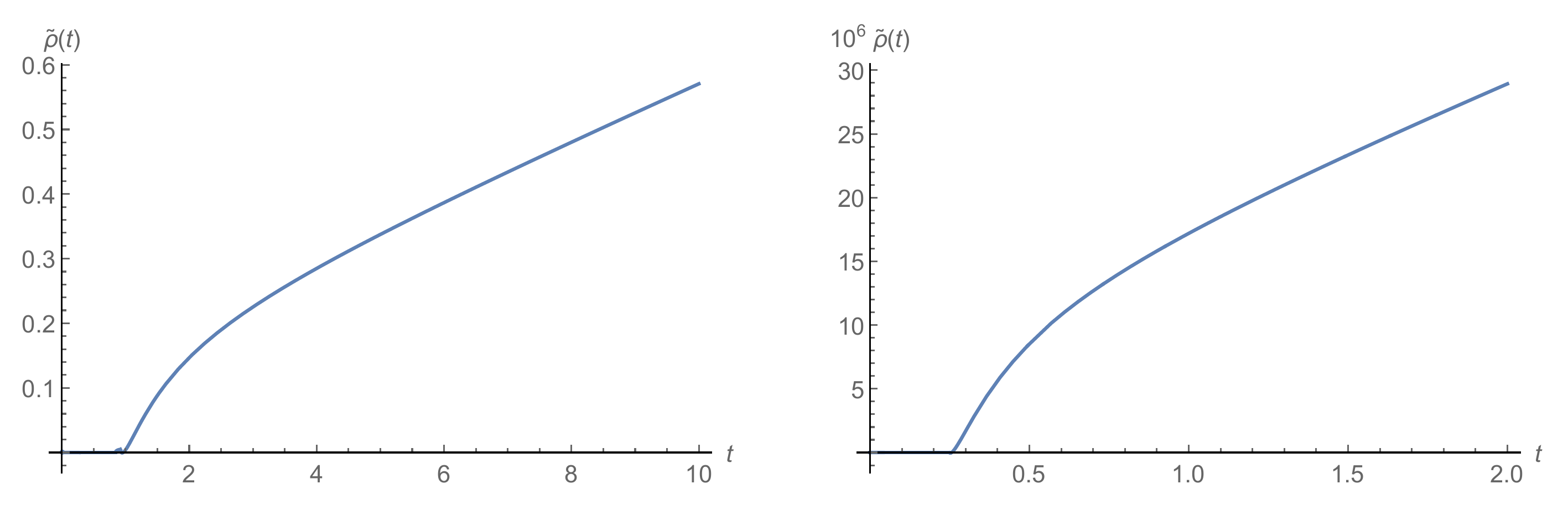}
			\caption{Spectral function for the propagator $\langle A^a_{\mu}(p) A^b_{\nu}(-p) \rangle$ in the unitary gauge, with $t$ given in unity of $\mu^2$, for  the Region I (left) and Region II (right), with parameter values given in Table \ref{tabel}.}
			\label{Yt}
		\end{figure}

Nevertheless, it is interesting to observe that, if we remove the divergent  part \eqref{divu} anyway,  we obtain the spectral function as shown in  FIG.~\ref{Yhh}. This spectral function is almost identical to that obtained for the composite operator $O(x)$, see FIG.~\ref{Y}.
	For the inverse gauge field propagator, proceeding in the same way, we find the divergent part
		\beq
		G^{-1}_{AA, \rm div}(p^2)&=&\frac{1}{\epsilon }\Big(-\frac{9 g^2 m^4}{16 \pi ^2 m_h^2}-\frac{g^2 p^6}{96 \pi ^2 m^4}+\frac{7 g^2 p^4}{48 \pi ^2 m^2}+\frac{3 g^2 m^2}{32 \pi ^2}+\frac{83 g^2 p^2}{96 \pi ^2}-\frac{3 \lambda  m^2}{8 \pi ^2}\Big) \label{gdivu}
		\eeq
		which shows again the non-renormalizability of unitary gauge,  through the terms $\sim \frac{p^6}{\epsilon m^4}$ and $\sim \frac{p^4}{\epsilon m^2}$ . However, if we remove also here those terms by hand,   we obtain the spectral function as shown in  FIG.~\ref{Yt}.

Apart from the by hand removal of the divergences in this unitary gauge exercise, the nice behaviour of the spectral densities for the Higgs and gauge field\footnote{Notice that these dropped divergences do not contribute anyhow to the branch cut discontinuity and this spectral function.} obtained by a direct use of the tree level propagators already taken in the unitary limit, $\xi \rightarrow \infty$, can be, to some extent, justified by the fact that we are working at the one-loop order in perturbation theory. Since overlapping divergences start from two-loop onward, we can easily figure out that the naive use of the elementary tree level propagators taken already in the unitary limit will run into  severe non-renormalizibility issues, making the removal of the  (overlapping) divergent parts \eqref{divu}, \eqref{gdivu} quite problematic beyond the current one-loop level.

\begin{figure}[H]
		\centering
		\includegraphics[width=\linewidth]{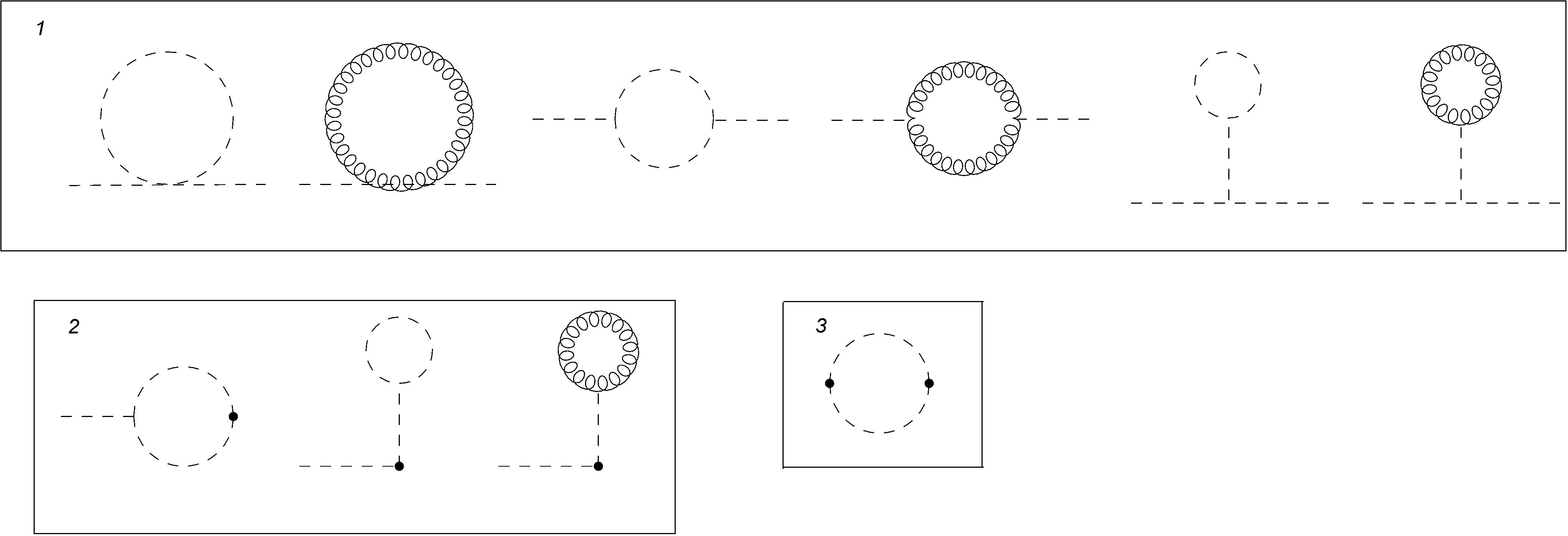}
		\caption{One-loop contributions for the propagator $\langle O(x)O(y) \rangle$ in the unitary gauge. Curly lines represent the gauge field and dashed lines the Higgs field. The $\bullet$  indicates the insertion of a composite operator.}.
		\label{OOu}
	\end{figure}

Now for what concerns the composite operators, since $O(x)$ is  BRST invariant, any choice for the gauge fixing should give the same expression for the correlation function  $G_{OO}(p^2)$. So again using the unitary gauge, eq.~\eqref{exp1} simplifies to
	\beq
	\braket{{O}(x) {O}(y)}_{\rm unitary} &=& v ^2 \braket{h(x) h(y)}_{\rm unitary}+ v  \braket{h(x) h(y)^2}_{\rm unitary}+\frac{1}{4}\braket{h(x)^2 h(y)^2}_{\rm unitary},
	\eeq
	with the consecutive terms displayed diagrammatically by the respective boxes in FIG.~\ref{OOu}. Using dimensional regularization in the  $\MSbar$-scheme with $(d=4-\epsilon)$ and switching to momentum space, we find	\beq
	v^2 \braket{h(p)h(-p)}_{\rm unitary}&=&\frac{3}{32 \pi^2  }\int_0^1 dx\Bigg\{\frac{1}{ \epsilon} \left(2m_h^4+12 m^2 p^2+2p^4\right)+2m_h^4  \ln \left(\frac{m_h^2}{\mu ^2}\right)+2 \left(6 m^4-m^2 p^2\right) \ln \left(\frac{m^2}{\mu ^2}\right)\nonumber\\
	&-&3 m_h^4 \ln \left(\frac{p^2 (1-x) x+m_h^2}{\mu ^2}\right)- \left(12 m^4+4 m^2 p^2+p^4\right) \ln \left(\frac{p^2 (1-x) x+m^2}{\mu ^2}\right)\nonumber\\
	&-&2m_h^4 +2 m^2 \left(p^2-6 m^2\right)\Bigg\}\frac{1}{\left(m_h^2+p^2\right){}^2},\\
	v \braket{h(p) h(-p)^2}_{\rm unitary}&=& \frac{3}{16 \pi ^2 m_h^2 }\int_0^1 dx \Bigg\{\frac{12}{\epsilon}m^4-6 m^4\ln \left(\frac{m^2}{\mu ^2}\right)-m_h^4\ln \left(\frac{m_h^2}{\mu ^2}\right)\nonumber\\
	&+&  m_h^4 \ln \left(\frac{m_h^2+p^2 (1-x) x}{\mu ^2}\right)+m_h^4+2m^4 \Bigg\}\frac{1}{\left(m_h^2+p^2\right)},\\
	\frac{1}{4}\braket{h(p)^2 h(p)^2}_{\rm unitary}&=& \frac{1}{16 \pi ^2  }\int_0^1 dx \Bigg\{\frac{1}{\epsilon}-\frac{1}{2}\ln \left(\frac{m_h^2+p^2 (1-x) x}{\mu ^2}\right)\Bigg\}.
	\eeq
	Inserting now the unity
	\beq
	1= (p^2+m_h^2)/(p^2+m_h^2) = ((p^2+m_h^2)/(p^2+m_h^2))^2,
	\label{2o}
	\eeq
	we find indeed that the finite pieces of $\braket{O(x)O(y)}_{\rm unitary}$ and $\braket{O(x)O(y)}_{R_\xi}$ do coincide, this is evidently related to the gauge invariance. From two-loop onward the overlapping divergences will show up again, requiring a fully renormalizable setup.

\begin{figure}[H]
		\centering
		\includegraphics[width=\linewidth]{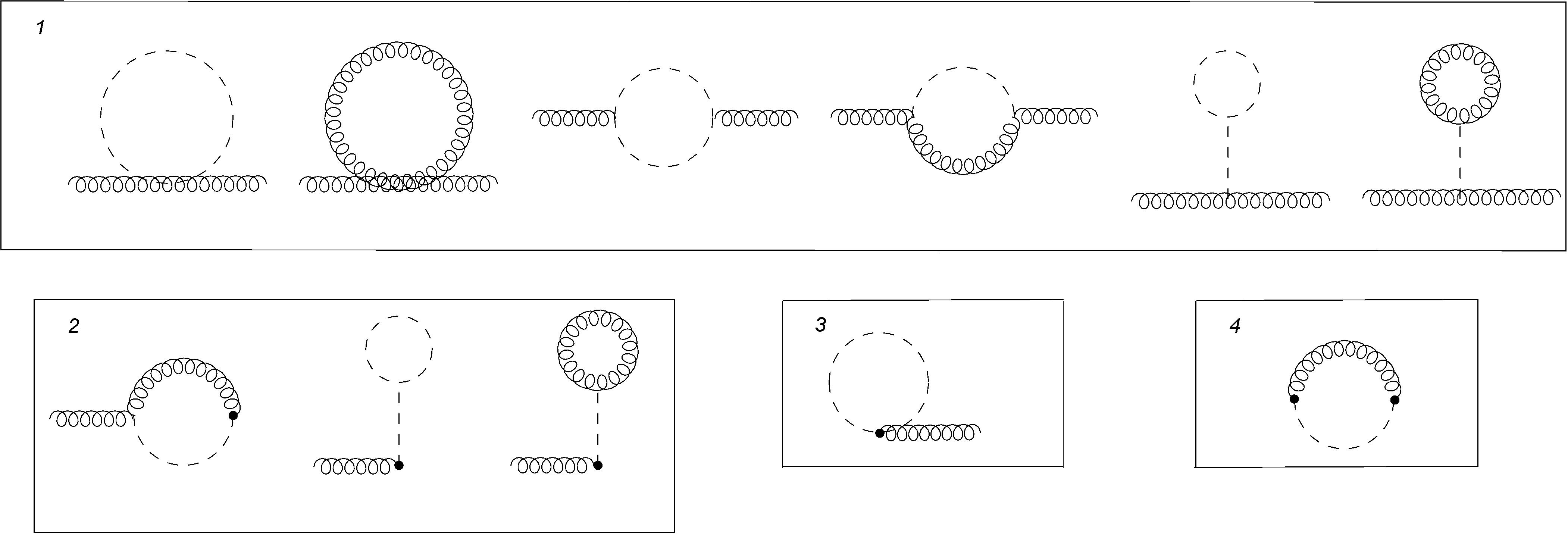}
		\caption{One-loop contributions for the propagator $\langle R^a_{\mu}(x)R^a_{\nu}(y) \rangle$ in the unitary gauge. Curly lines represent the gauge field and dashed lines the Higgs field. The $\bullet$  indicates the insertion of a composite operator.}
		\label{RRu}
	\end{figure}

We can stretch the unitary gauge even a bit further, to also look at the vector correlator, $\braket{R^a_{\mu}(p)\,R^a_{\nu}(-p)}_{\rm unitary}$. We find at one-loop order
	\beq
	\braket{R^a_{\mu}(p)\,R^a_{\nu}(-p)}&=&\frac{1}{16}g^{2}v^{4}\langle A_{\mu}^{a}(p)\,A_{\nu}^{a}(-p)\rangle_{\rm unitary}+\frac{1}{4}g^{2}v^{3}\langle A_{\mu}^{a}(p)\,(A_{\nu}^{a}h)(-p)\rangle_{\rm unitary}\nonumber\\
	&+&\frac{1}{8}g^{2}v^{2}\langle (h^2 A_{\mu}^{a})(p)\,A_{\nu}^{a}(-p)\rangle_{\rm unitary}+\frac{1}{4}g^{2}v^{2}\langle (h A_{\mu}^{a})(p)\,(h A_{\nu}^{a})(-p)\rangle_{\rm unitary}
	\eeq
	with the consecutive terms displayed diagrammatically by the respective boxes  in FIG.~\ref{RRu}.

	 We find in momentum space
	\beq
	\frac{1}{16}g^{2}v^{4}\langle A_{\mu}^{a}(p)\,A_{\nu}^{a}(-p)\rangle_{\rm unitary}&=& \frac{g^4 v^4}{32 (4\pi)^2} \int_0^1 dx\Bigg\{ -\frac{1}{\epsilon}\frac{1}{6m^4 m_h^2}\left(9 m^4 m_h^4+m_h^2 \left(-9 m^6-83 m^4 p^2-14 m^2 p^4+p^6\right)+54 m^8\right)\nonumber\\
	&+&\frac{m_h^2}{2 p^2} \left(-m_h^2+m^2+7 p^2\right) \ln \left(\frac{m_h^2}{\mu ^2}\right)\nonumber\\
	&+&\frac{1}{2 m^2 p^2 m_h^2}\left(m^4 m_h^4-m_h^2 \left(m^6+47 m^4 p^2+16 m^2 p^4-2 p^6\right)+54 m^6 p^2\right) \ln \left(\frac{m^2}{\mu ^2}\right)\nonumber\\
	&+&\frac{1}{2p^2}\left(-2 m_h^2 \left(m^2-p^2\right)+m_h^4+m^4-10 m^2 p^2+p^4\right)\ln \left(\frac{p^2 (1-x) x+(1-x) m_h^2+m^2 x}{\mu ^2}\right)\nonumber\\
	&+&\frac{1}{2 m^4}\left(4 m^2+p^2\right) \left(12 m^4-20 m^2 p^2+p^4\right) \ln \left(\frac{m^2+p^2 (1-x) x}{\mu ^2}\right)\nonumber\\
	&+&\frac{1}{6 m^4 p^2 m_h^2}\Big(3 m^4 m_h^6-3 m_h^4 \left(2 m^6+9 m^4 p^2\right)\nonumber\\
	&+&m_h^2 \left(3 m^8-9 m^6 p^2-2 m^4 p^4-26 m^2 p^6-2 p^8\right)-54 m^8 p^2\Big)\Bigg\}\frac{\mathcal{P}_{\mu \nu}(p)}{(p^2+m^2)^2}\nonumber\\
	&+& \frac{g^4 v^4}{32 (4\pi)^2m^4} \int_0^1 dx\Bigg\{\frac{1}{\epsilon}\frac{3}{ m_h^2} \left(m_h^2 \left(3 m^2+p^2\right)-3 m_h^4-18 m^4\right)\nonumber\\
	&-&\frac{3 m^2}{2  p^2 m_h^2}\left(m_h^2 \left(p^2-m^2\right)+m_h^4-18 m^2 p^2\right) \ln \left(\frac{m^2}{\mu ^2}\right)\nonumber\\
	&+&\frac{3 m_h^2}{2 p^2} \left(m_h^2-m^2+5 p^2\right) \ln \left(\frac{m_h^2}{\mu ^2}\right)\nonumber\\
	&-&\frac{3}{2 p^2} \left(\left(m_h-m\right){}^2+p^2\right) \left(\left(m_h+m\right){}^2+p^2\right) \ln \left(\frac{p^2 (1-x) x+(1-x) m_h^2+m^2 x}{\mu ^2}\right)\nonumber\\
	&-&\frac{3}{2p^2 m_h^2} \left(m_h^4 \left(5 p^2-2 m^2\right)+m_h^2 \left(m^4-m^2 p^2\right)+m_h^6+6 m^4 p^2\right)\Bigg\} \mathcal{L}_{\mu \nu}(p),\nonumber\\
	\frac{1}{4}g^2 v^3 \braket{A^a_{\mu}(p)\,(A_\nu^a h)(-p)}_{\rm unitary}&=&\frac{1}{16\pi ^2 }\int_0^1 dx \Bigg\{\frac{1}{\epsilon}\frac{m^2}{ m_h^2} \left(m_h^2 \left(p^2-9 m^2\right)+12 m_h^4+54 m^4\right)\nonumber\\
	&-&\frac{m^2 m_h^2}{2 p^2} \left(-m_h^2+m^2+10 p^2\right) \ln \left(\frac{m_h^2}{\mu ^2}\right)\nonumber\\
	&-&\frac{m^4}{2 p^2 m_h^2} \left(m_h^2 \left(p^2-m^2\right)+m_h^4+54 m^2 p^2\right) \ln \left(\frac{m^2}{\mu ^2}\right)\nonumber\\
	&-&\frac{m^2}{2p^2} \left(2 p^2 \left(m_h^2-5 m^2\right)+\left(m^2-m_h^2\right){}^2+p^4\right) \ln \left(\frac{p^2(1-x)x+(1-x)m_h^2+x m^2}{\mu ^2}\right)\nonumber\\
	&+&\frac{m^2}{6 p^2 m_h^2} \left(6 m_h^4 \left(m^2+6 p^2\right)+m_h^2 \left(-3 m^4+9 m^2 p^2+2 p^4\right)-3 m_h^6+54 m^4 p^2\right)\nonumber\Bigg\}\frac{\mathcal{P}_{\mu \nu}(p)}{\left(m^2+p^2\right)}\nonumber\\
	&+&\frac{1}{16\pi ^2}\int_0^1 dx \Bigg\{
	\frac{1}{\epsilon}\frac{3}{  m_h^2} \left(-m_h^2 \left(3 m^2+p^2\right)+m_h^4+18 m^4\right)\nonumber\\
	&+&\frac{3 m^2}{2 p^2 m_h^2} \left(m_h^2 \left(p^2-m^2\right)+m_h^4-18 m^2 p^2\right) \ln \left(\frac{m^2}{\mu ^2}\right)\nonumber\\
	&-&\frac{3 m_h^2}{2 p^2} \left(m_h^2-m^2+3 p^2\right) \ln \left(\frac{m_h^2}{\mu ^2}\right)\nonumber\\
	&+&\frac{3}{2 p^2 m_h^2} \left(m_h^2 \left(\left(m-m_h\right){}^2+p^2\right) \left(\left(m_h+m\right){}^2+p^2\right)\right) \ln \left(\frac{p^2(1-x)x+(1-x)m_h^2+ x m^2}{\mu ^2}\right)\nonumber\\
	&+&\frac{3}{2 p^2 m_h^2} \left(\left(m_h^3-m^2 m_h\right){}^2+p^2 \left(-m^2 m_h^2+3 m_h^4+6 m^4\right)\right)\Bigg\}\mathcal{L}_{\mu \nu }(p),
	\eeq
	\beq
	\frac{1}{8}g^{2}v^{2}\langle (h^2 A_{\mu}^{a})(p)\,A_{\nu}^{a}(-p)\rangle_{\rm unitary}&=& -\frac{3 m_h^2 m^2}{32\pi^2}\int_0^1 dx \Bigg\{\frac{2}{\epsilon }- \ln \left(\frac{m_h^2}{\mu ^2}\right)+1\Bigg\}\Big(\frac{1}{p^2+m^2}\mathcal{P}_{\mu \nu}(p)+\frac{1}{m^2}\mathcal{L}_{\mu\nu}(p)\Big),
	\eeq
	\beq
	\frac{1}{4}g^{2}v^{2}\langle (h A_{\mu}^{a})(p)\,(h A_{\nu}^{a})(-p)\rangle_{\rm unitary}&=&\frac{1}{32\pi ^2} \int_0^1 dx \Bigg\{\frac{1}{ \epsilon }(-3 m_h^2+9 m^2-p^2)\nonumber\\
	&+&\frac{m_h^2 }{2 p^2}\left(m^2+p^2-m_h^2\right) \ln \left(\frac{m_h^2}{\mu ^2}\right)+\frac{m^2 }{2 p^2} \left(m_h^2-m^2+p^2\right)\ln\left(\frac{m^2}{\mu ^2}\right)\nonumber\\
	&+&\frac{1}{2p^2} \left(2 p^2 \left(m_h^2-5 m^2\right)+\left(m^2-m_h^2\right){}^2+p^4\right) \ln \left(\frac{p^2(1-x)x+(1-x)m_h^2+x m^2}{\mu ^2}\right)\nonumber\\
	&+&\frac{1}{6p^2} \left(3 \left(m^2-m_h^2\right){}^2-9 p^2\left(m_h^2+m^2\right)-2 p^4\right)\Bigg\}\mathcal{P}_{\mu \nu}(p)\nonumber\\
	&+& \frac{3}{32\pi ^2}\int_0^1 dx \Bigg\{\frac{1}{ \epsilon } \left(-m_h^2+3 m^2+p^2\right)\nonumber\\
	&+&\frac{m^2}{2 p^2} \left(-m_h^2+m^2-p^2\right) \ln \left(\frac{m^2}{\mu ^2}\right)+\frac{m_h^2}{2 p^2} \left(m_h^2-m^2+3 p^2\right) \ln \left(\frac{m_h^2}{\mu ^2}\right)\nonumber\\
	&-& \frac{1}{2p^2} \left(\left(m-m_h\right){}^2+p^2\right) \left(\left(m_h+m\right){}^2+p^2\right) \ln \left(\frac{p^2x(1-x)+(1-x)m_h^2+x m^2}{\mu ^2}\right)\nonumber\\
	&-&\frac{1}{2p^2}\left(\left(m^2-m_h^2\right){}^2+3 p ^2m_h^2-p^2m^2\right)\Bigg\}\mathcal{L}_{\mu \nu}(p) \;.
	\eeq
and putting everything together, also here we see that $\braket{R^a_{\mu}(p)\,R^a_{\nu}(-p)}_{\rm unitary}=\braket{R^a_{\mu}(p)\,R^a_{\nu}(-p)}_{R_\xi}$ upon dropping the divergences.		
	\end{appendix}

	\bibliographystyle{apsrev4-1}
	
	\bibliography{ref2}
	
\end{document}